\documentclass[sort&compress]{jfm}

\usepackage{amsmath,amssymb,mathtools}
\usepackage{graphicx}% Include figure files
\usepackage{dcolumn}% Align table columns on decimal point
\usepackage{bm}% bold math
\usepackage{hyperref}% add hypertext capabilities
\usepackage{siunitx}
\usepackage[noabbrev,nameinlink]{cleveref}
\usepackage[labelformat=simple]{subcaption}

\usepackage{overpic}

\usepackage[normalem]{ulem}
\usepackage{comment}
\usepackage[usenames,dvipsnames]{color}
\usepackage{natbib}

\newcommand{\diff}[2]{\frac{\mathrm{d}#1}{\mathrm{d}#2}}
\renewcommand{\vec}[1]{\bm{#1}}
\newcommand{\abs}[1]{\left\lvert #1\right\rvert}
\newcommand{\norm}[1]{\left\lVert #1\right\rVert}
\newcommand{\intd}[1]{~\mathrm{d}#1}
\newcommand{\cross}{\mathbin{\times}}

\newcommand{\tensor}[1]{\mathsfbi{#1}}

\newcommand{\ds}{\Delta \mathrm{s}}
\DeclareMathOperator{\sgn}{sgn}
\newcommand{\ex}{\vec{e}_x}
\newcommand{\ey}{\vec{e}_y}
\newcommand{\ez}{\vec{e}_z}

\shorttitle{Regularised non-uniform segments and efficient no-slip elastohydrodynamics}
\shortauthor{B. J. Walker and E. A. Gaffney}
\title{Regularised non-uniform segments and efficient no-slip elastohydrodynamics}

\author{B. J. Walker\aff{1\corresp{\email{benjamin.walker@maths.ox.ac.uk}}} \and E. A. Gaffney\aff{1}}

\affiliation{\aff{1}Wolfson Centre for Mathematical Biology, Mathematical Institute, University of Oxford, Oxford, OX2 6GG, UK}

\begin{document}

\maketitle

\begin{abstract}
%auto-ignore
The elastohydrodynamics of slender bodies in a viscous fluid have long been
the source of theoretical investigation, being pertinent to the microscale
world of ciliates and flagellates as well as to biological and engineered
active matter more generally. Though recent works have overcome the severe
numerical stiffness typically associated with slender elastohydrodynamics,
employing both local and non-local couplings to the surrounding fluid, there
is no framework of comparable efficiency that rigorously justifies its
hydrodynamic accuracy. In this study, we combine developments in filament
elastohydrodynamics with a recent slender-body theory, affording algebraic
asymptotic accuracy to the commonly imposed no-slip condition on the surface
of a slender filament of potentially non-uniform cross-sectional radius.
Further, we do this whilst retaining the remarkable practical efficiency of
contemporary elastohydrodynamic approaches, having drawn inspiration from the
method of regularised Stokeslet segments to yield an efficient and flexible
slender-body theory of regularised non-uniform segments.
\end{abstract}

% Main text.
\section{Introduction}
\label{sec:intro}
%autoignore
% General background of filaments.
The coupled elastohydrodynamics of flexible slender filaments are of intense
interest to a breadth of active research communities, ranging from theoretical
to experimental studies of filaments from the perspectives of synthetic
sensors to those rooted in the biology and mechanics of cilia and flagella
\citep{Pozrikidis2010,Gray1928,Smith2019,Roper2006,Guglielmini2012,Curtis2012,Simons2015}.
A comprehensive summary of the field is given in the recent review of
\citet{DuRoure2019}, which notes a particular need for further theoretical
development in this area. Indeed, up until recently, problems involving
filament elastohydrodynamics have been largely out of reach due to severe
numerical stiffness associated with the dynamics of a slender body in a
viscous fluid, with few studies being able to utilise large computing
resources to combat this issue \citep{Olson2013,Ishimoto2018a,Schoeller2018}.
However, the work of \citet{Moreau2018} sought to address such problems,
integrating the governing equations of elasticity in space in order to
generate a coarse-grained framework with greatly reduced numerical stiffness.
Despite being a recent development in the field, this approach has already
been extended by \citet{Hall-McNair2019} and \citet{Walker2019d} to include
improved non-local hydrodynamics, applied to the model biological problem of
flagellar efficiency \citep{Neal2020}, and extended to motion in three
dimensions \citep{Walker2019g}.

% Simple hydrodynamics via ansatz lacking proper justification.
Common to these recent models, as well as to other treatments of slender
filaments at zero Reynolds number, are simplified representations of
slender-body hydrodynamics. The aforementioned work of \citet{Moreau2018}
utilises resistive force theory, a local relation between motion and drag that
has seen widespread use since its advent in the 1950s
\citep{Hancock1953,Gray1955}. More refined and complex are slender-body
theories, which capture the non-local coupling of kinematics and associated
forces via an integral relation, as considered in the early studies of
\citet{keller1976,Cox1970,lighthill1976} and later refined by
\citet{Johnson1980}. Use of these slender theories in numerical applications
often necessitates the use of many-point quadrature rules or specialised
techniques to evaluate the integral of a rapidly varying or singular kernel,
issues also found in methods derived from the boundary integral formulation of
Stokes equations, as summarised by \citet{pozrikidis1992}. In the early 2000s,
\citet{Cortez2001} circumvented such issues of numerical complexity by instead
considering solutions of the regularly forced Stokes equations, leading to a
regularised Green's function and an associated regularised theory. In turn,
drawing from significant earlier study of singular slender-body theories, this
led to commonplace use of a regularised slender-body theory ansatz for flow
around a slender filament in terms of a force density $\vec{f}$, typically an
integral over the centreline of the filament of the form
\begin{equation}\label{eq:general_ansatz}
	\vec{u}(\vec{x}) = \int \tensor{K}^{\epsilon}(\vec{x},s')\vec{f}(s')\intd{s'}\,,
\end{equation}
where $\vec{u}(\vec{x})$ is the fluid velocity at a point $\vec{x}$ and
$\tensor{K}^{\epsilon}$ is a regular integral kernel. The parameter $\epsilon$
represents a lengthscale of the regularisation, which in studies of filament
dynamics has often been taken to be the filament radius without rigorous
justification
\citep{Cortez2012,Walker2019d,Hall-McNair2019,Cortez2018,Smith2009d}, with
circular cross sections invariably assumed. The general ansatz of
\cref{eq:general_ansatz} is also commonly used in conjunction with the
hydrodynamic no-slip condition, though is evaluated not on the surface of the
body, but on the filament centreline. With many approaches taking the integral
kernel $\tensor{K}^{\epsilon}$ to simply be the regularised point force
Green's function in the appropriate domain, application of this approximate
relation does not guarantee that the no-slip boundary condition is satisfied
on the surface of the body, with particular issues arising at the endpoints of
the flagellum, where more than a velocity Green's function can be required
\citep{Chwang1975}.

Building upon the singular work of \citet{Johnson1980} and the classical
solution of \citet{Chwang1975} for a prolate ellipsoid, the recent theory of
\citet{Walker2020b} surpasses these general shortfalls and leverages a
particular choice of kernel $\tensor{K}^{\chi}$, along with a systematically
justified and spatially dependent regularisation parameter $\chi$, to satisfy
the no-slip boundary condition on the surface of a slender body up to errors
algebraic in the body aspect ratio. This theory retains the non-singular
nature and accompanying numerical simplicity of the general regularised
ansatz, whilst affording systematically justified accuracy and
parameterisation. With such features having been absent from the recent
efficient frameworks of \citet{Moreau2018,Hall-McNair2019,Walker2019d}, the
primary aim of this study is to incorporate the theory of \citet{Walker2020b}
into the coarse-grained elastohydrodynamic framework of \citet{Walker2019d},
enabling the efficient simulation of slender bodies with asymptotically
justified hydrodynamic accuracy in the no-slip condition. In doing so, we will
additionally attempt to address concerning oscillations present in the force
density solutions of these frameworks, which reportedly persist even with
improved filament discretisations \citep{Walker2019d,Cortez2018}.

However, whilst the incorporation of the simple ansatz of \citet{Walker2020b}
may be achieved with relative ease, integration of the regular but rapidly
varying kernels may limit the speed of computation if performed with
quadrature, as implemented in the original work of \citet{Walker2020b}. Having
built upon the works of \citet{Smith2009d} and \citet{Cortez2018},
respectively, \citet{Hall-McNair2019} and \citet{Walker2019d} avoid such
expensive computation by analytically integrating the kernel over the straight
line segments that form the discretised centreline of the slender body, which
we will refer to as the regularised Stokeslet segment (RSS) approach. Though
complicated here by a non-constant regularisation parameter $\chi$, we will
aim to proceed in a similar fashion and remove the reliance on quadrature
rules in order to realise a highly efficient numerical framework for the study
of slender-body elastohydrodynamics.

% Summary
Hence, we will proceed by first defining the non-uniform filament problem,
adopting and unifying the notation of \citet{Walker2020b} and
\citet{Walker2019d} for slender-body kinematics. We then describe a
modification of the coarse-grained framework of \citet{Moreau2018}, similar in
form to that of \citet{Walker2019d}, and present the slender-body theory of
\citet{Walker2020b} cast in dimensionless quantities. Having adopted a
piecewise-constant discretisation of viscous force density, we then seek to
perform the slender-body integrals analytically, Taylor expanding the
regularisation parameter $\chi$ to yield symbolic tractability. We will then
numerically evidence the improved satisfaction of the no-slip boundary
condition on the surface of the filament attained with the presented
methodology, in turn considering the computed profiles of force density along
the centreline of the filament and their behaviour near the endpoints of the
slender body.
\section{The non-uniform filament problem}
\label{sec:setup}
\begin{figure}
	\centering
	\begin{subfigure}[c]{0.49\textwidth}
		\centering
		\vspace{1.01em}
		\includegraphics[height=0.4\textwidth]{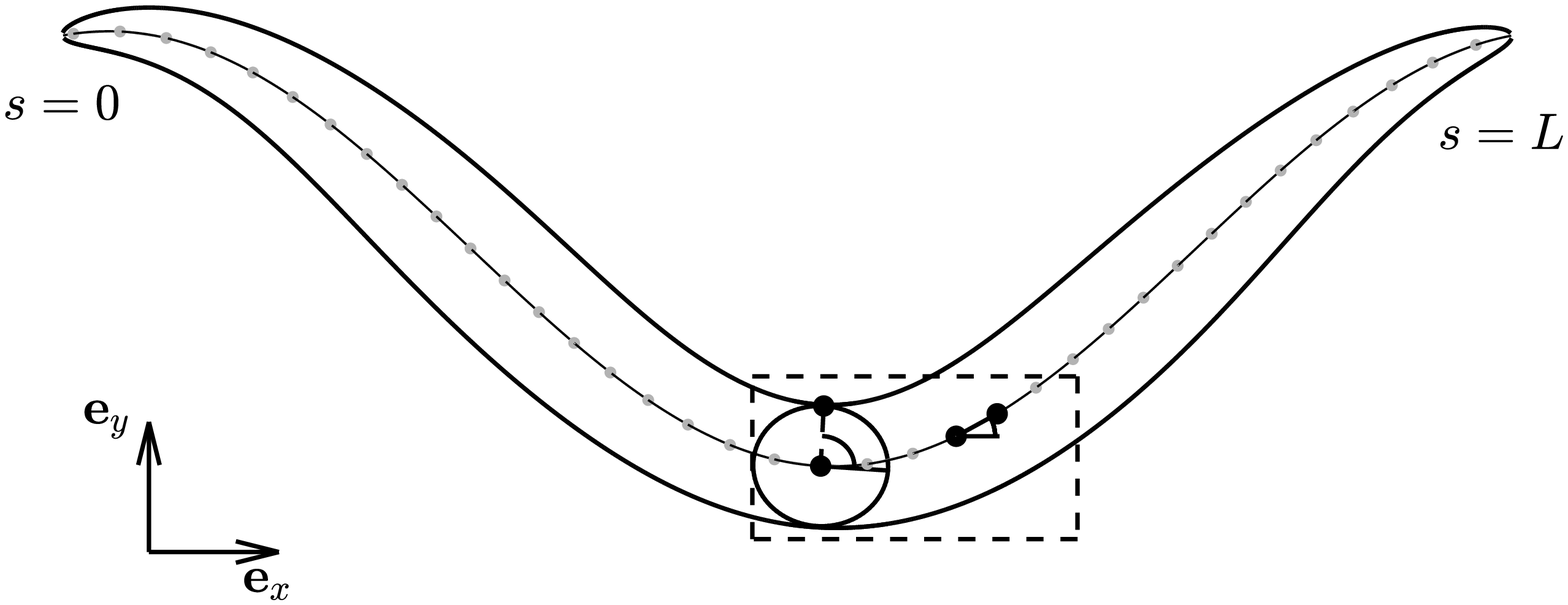}
		\caption{\label{fig:def:setup}}
	\end{subfigure}\ \
	\begin{subfigure}[c]{0.49\textwidth}
		\centering
		\vspace{1.01em}
		\includegraphics[height=0.4\textwidth]{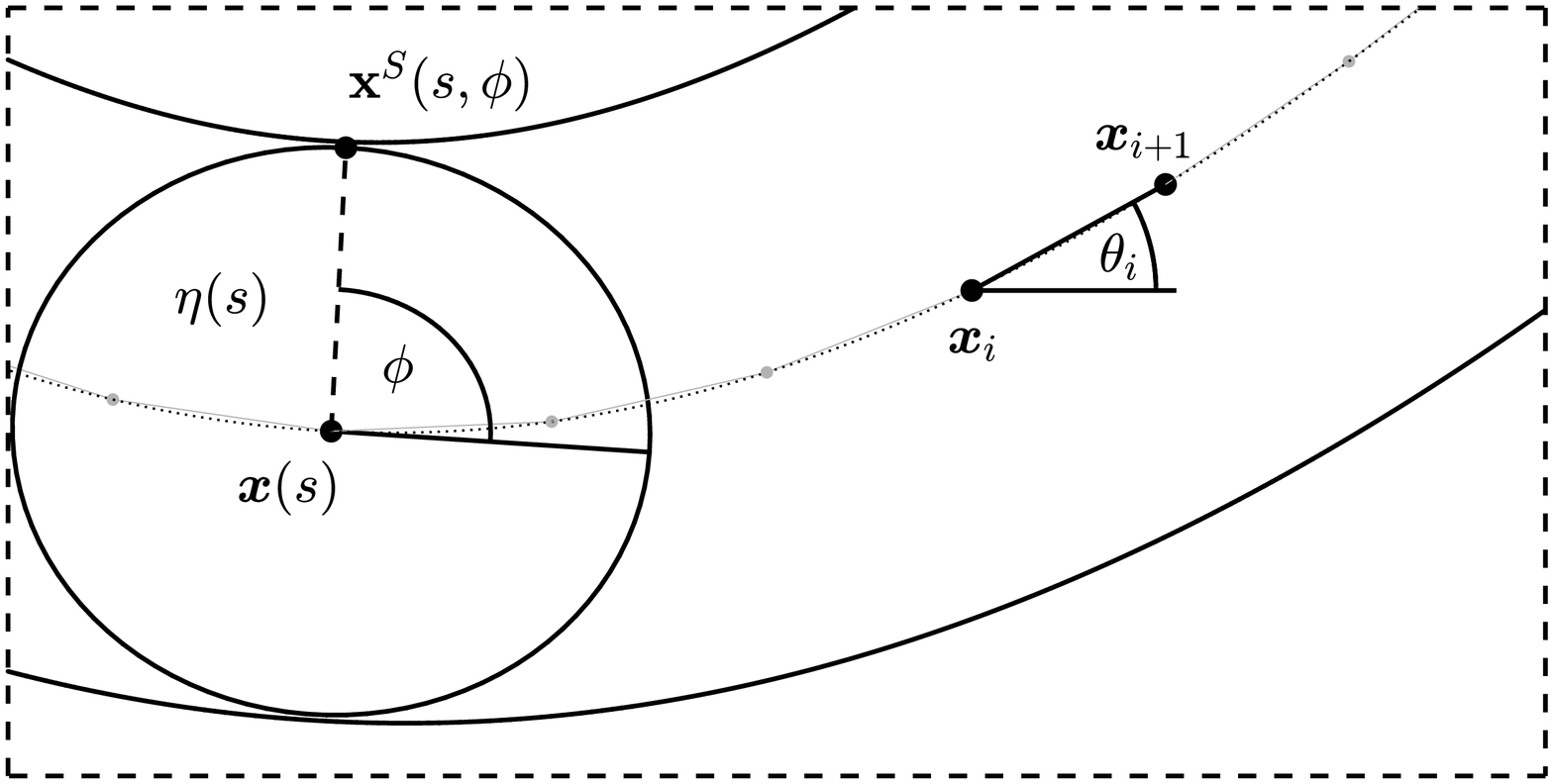}
		\caption{\label{fig:def:setup_zoomed}}
	\end{subfigure}
	\caption{Filament setup and notation. (a) A general locally axisymmetric
	filament of total length $L$, with its centreline contained in a plane
	spanned by $\ex$ and $\ey$. (b) A zoomed view of the slender body, with
	centreline $\vec{x}(s)$ and associated surface points $\vec{x}^S(s,\phi)$
	parameterised by angle $\phi$ at a distance $\eta(s)$ from the centreline.
	Discrete points are shown as grey circles, connected by solid straight
	line segments that approximate the continuous dotted centreline. Example
	such discrete points $\vec{x}_i$ and $\vec{x}_{i+1}$ are highlighted in
	black, with the connecting line segment defining the angle $\theta_i$
	relative to the fixed $\vec{e}_x$ direction.}
	\label{fig:def}
\end{figure}

%autoignore
In this work we will consider the planar motions of a thin inextensible,
unshearable, untwistable filament in a viscous fluid, with the filament
centreline denoted $\vec{x}(s,t)= x(s,t)\ex+y(s,t)\ey$, without loss of
generality, where $\ex,\ey$ are constant orthogonal unit vectors in a fixed
inertial reference frame and span the plane of motion. Here, $s\in[0,L]$ is an
arclength parameter and time is denoted by $t$, where $L$ is the length of the
slender object. Distinct from the notation of the Introduction, this
slenderness is captured by the dimensionless parameter $\epsilon$, defined
explicitly as
\begin{equation}
	\epsilon = \frac{2\max_{s\in[0,L]}\{\eta(s)\}}{L}\ll1\,,
\end{equation}
where $\eta(s)$ is the non-negative radius of the filament at arclength $s$,
having assumed local axisymmetry about the centreline. With the shape
therefore entirely defined by the centreline and radius function, we may
describe points on the surface of the filament as
\begin{equation}
	\vec{x}^S(s,\phi) = \vec{x}(s) + \eta(s)\vec{e}_r(s,\phi)\,,
\end{equation}
where $\phi$ is a cross-sectional angle. Here, $\vec{e}_r$ is a radial unit
vector embedded in a transverse cross section to the centreline. For unit
tangent, normal, and binormal unit vectors defined by the Frenet-Serret
relations
\begin{equation}
	\vec{e}_t(s) = \frac{\partial \vec{x}}{\partial s}, \qquad \frac{\partial \vec{e}_t}{\partial s} = \theta_s \vec{e}_n(s), \qquad
	\vec{e}_b(s) = \vec{e}_t(s) \cross \vec{e}_n(s)\,,
\end{equation}
where $\theta(s,t)$ defines the filament tangent angle relative to $\ex$, we
define
\begin{equation}
	\vec{e}_r(s,\phi) = \vec{e}_n(s) \cos \phi + \vec{e}_b(s) \sin \phi\,.
\end{equation}
Here and throughout, subscripts of $s$ denote derivatives with respect to
arclength and we have omitted writing the inherent time dependence of the
filament centreline and all derived quantities. These definitions are
illustrated in \cref{fig:def}.

We discretise the filament centreline into $N$ linear segments, with the
endpoints of these segments denoted by $\vec{x}(s_i)$ for uniformly spaced
arclengths $s_i=(i-1)L/N \in[0,L]$, where $i=1,\ldots,N+1$. We write
$\vec{t}_i$ for the unit tangent to each linear segment, noting that this is
an approximation of $\vec{e}_t(s)$ on the $i$th segment, and parameterise
these discrete tangents by $\theta(s)$, itself discretised as
$\theta(s)\approx\theta_i$ on the $i$th segment such that $\vec{t}_i =
\cos{\theta_i}\ex + \sin{\theta_i}\ey$. With this piecewise linear
discretisation of $\vec{x}$ in arclength, or equivalently a piecewise constant
discretisation of $\theta$, we may describe the position of the filament with
only the $N+2$ quantities $x_1,y_1,\theta_1,\ldots,\theta_N$, where $\vec{x}_1
= x_1\ex+y_1\ey$. Explicitly, for $j=1,\ldots,N+1$ we have
\begin{equation}
	\vec{x}_j = \vec{x}_1 + \sum\limits_{i=1}^{j-1}(\cos{\theta_i}\ex + \sin{\theta_i}\ey)\ds\,,
\end{equation}
where $\ds$ is the constant segment length, equivalently defined as $\ds =
L/N$. Differentiating with respect to time, denoting time derivatives with a
dot, this gives the linear velocity of the material point $\vec{x}_j$ as
\begin{equation}\label{eq:endpoint_vel}
	\dot{\vec{x}}_j = \dot{\vec{x}}_1 + \sum\limits_{i=1}^{j-1}(-\sin{\theta_i}\ex + \cos{\theta_i}\ey)\dot{\theta}_i\ds\,.
\end{equation}
We may concisely write this latter linear relation as
\begin{equation}\label{eq:QTX}
	\tensor{Q}\dot{\vec{\theta}} = \dot{\vec{X}}\,,
\end{equation}
where $\vec{\theta} = [x_1,y_1,\theta_1,\ldots,\theta_N]^T$, $\vec{X} =
[x_1,y_1,\ldots,x_{N+1},y_{N+1}]^T$ and $\tensor{Q}$ is the linear
operator encoding \cref{eq:endpoint_vel}, the latter having dimension
$(2N+2)\times (N+2)$ and given explicitly in the work of \citet{Walker2019d}.
Hence, we may readily cast expressions involving $\dot{\vec{X}}$ in terms of
the reduced variables $\vec{\theta}$ and their time derivatives.

The equations governing the surrounding fluid medium will be the familiar
Newtonian Stokes equations, valid in the inertia-free limit of zero Reynolds
number, which we will assume throughout. This limit is relevant to a broad
range of biological and physical circumstances, for example the small-scale
beating of spermatozoan flagella or the bending of cilia in flow. The Stokes
equations may be briefly stated as
\begin{equation}\label{eq:stokes}
	\mu\nabla^2\vec{u} = \nabla p\,, \quad \nabla\cdot\vec{u}=0\,,
\end{equation}
where $\vec{u}$ is the fluid velocity, $\mu$ is the associated viscosity and
$p$ is the pressure. Here we will also assume that the flow is in an unbounded
domain in the exterior of the filament, and decays to zero in the far field.
\section{No-slip elastohydrodynamics}
\label{sec:method}
%auto-ignore
\subsection{Coarse-grained mechanics}\label{sec:coarse_graining}
Following \citet{Moreau2018}, we state the governing equations of
elasticity for this slender inextensible unshearable filament in pointwise
form as
\begin{align}
	\vec{n}_s - \vec{f} &= \vec{0}\,,\\
	\vec{m}_s + \vec{x}_s \cross \vec{n} &= \vec{0}\,,
\end{align}
for contact force and couple denoted $\vec{n},\vec{m}$ respectively and where
a subscript of $s$ denotes differentiation with respect to arclength. Here and
throughout, the filament is passive, with no driving internal couple and
$\vec{f}$ denotes the force per unit length applied on the surrounding fluid
by the filament. Note that the external couple exerted by the fluid on the
filament is $O(\epsilon^2)$, which will be negligible at the level of
asymptotic approximation that we will consider in this work. To proceed, we
integrate these equations with respect to arclength $s$, yielding
\begin{align}
	- \sum\limits_{j=1}^{N} \int\limits_{s_j}^{s_{j+1}} \vec{f}(s)\intd{s} &= \vec{n}(0)\,, \label{eq:integrated_balance_force}\\
	- \sum\limits_{j=i}^{N}\int\limits_{s_j}^{s_{j+1}} (\vec{x}(s)-\vec{x}_i)\cross\vec{f}(s)\intd{s} &= \vec{m}(s_i)\,, \quad i = 1,\ldots,N\,,\label{eq:integrated_balance_moment}
\end{align}
where we have decomposed the integrals into those over discrete segments and
integrated the pointwise moment balance from $s=s_i$ to $s=s_{N+1}=L$ for
$i=1,\ldots,N$. In writing
\cref{eq:integrated_balance_force,eq:integrated_balance_moment} we have
assumed that the filament is force and moment free at $s=L$, equivalent to
imposing $\vec{n}(L)=\vec{m}(L)=\vec{0}$. We additionally assume that these
conditions hold at the base, so that $\vec{n}(0) = \vec{m}(0) =
\vec{0}$, though each of these boundary conditions may be readily replaced
with those appropriate for particular problem settings, for example the
clamping of one end of the filament. Recalling that the considered filament
motion is purely planar, each term of \cref{eq:integrated_balance_moment} is
proportional to $\ex\cross\ey=\ez$, with $\vec{m}(s_i)=m(s_i)\ez$, so that
\cref{eq:integrated_balance_moment} collapses onto $N$ scalar equations. We
adopt a simple constitutive law, writing $\vec{m}(s_i)=EI\theta_s(s_i)\approx
EI(\theta_i-\theta_{i-1})/\ds$ for bending stiffness $EI$, valid for
$i=2,\ldots,N$.

\begin{figure}
	\centering
	\vspace{0.1em}
	\includegraphics[width=0.8\textwidth]{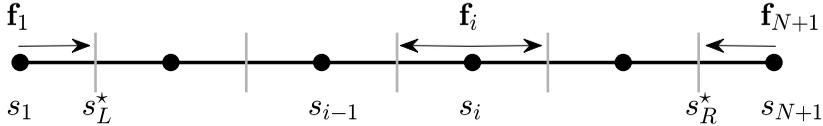}
	\caption{Illustration of the piecewise-constant force density
	discretisation. The horizontal line represents arclength $s$, with the
	discrete arclengths $s_i$ corresponding to segment endpoints shown as
	black circles. The force density $\vec{f}$ is approximated as taking the
	value $\vec{f}_i$ in a neighbourhood of the arclength $s_i$, typically
	between the midpoints $(s_{i-1}+s_i)/2$ and $(s_i + s_{i+1})/2$ of the
	adjacent segments, which are shown as vertical grey lines. The exceptional
	cases are on the first and final segments, where the midpoints are
	replaced with $s_L^{\star}$ and $s_R^{\star}$, respectively, in order to
	simplify the later description of the hydrodynamic slender-body theory.}
	\label{fig:force_discretisation}
\end{figure}

Illustrated in \cref{fig:force_discretisation}, we discretise the force
density $\vec{f}$, adopting a piecewise constant representation that is
distinct from that of $\theta$. Denoting the value taken by $\vec{f}$ at the
segment endpoints $\vec{x}_i$ by $\vec{f}_i$, for $i=1,\ldots,N+1$, we
discretise $\vec{f}$ as
\begin{equation}
	\vec{f}(s) = \left\{\begin{array}{ll}
		\vec{f}_i\,, & s\in[s_i,s_i+\frac{\ds}{2})\\
		\vec{f}_{i+1}\,, & s\in[s_i+\frac{\ds}{2},s_{i+1})\,,
	\end{array}\right.
\end{equation}
where $i\in\{2,\ldots,N-1\}$ is such that $s\in[s_i,s_{i+1})$. This is
equivalent to stating that, on segments $i=2,\ldots,N-1$, the value taken by
$\vec{f}$ is equal to that at the closest segment endpoint, with the $i$th
segment effectively split into two halves. The definition on the first and
last segments is similar, though the segment is not precisely split into two
equal parts, which will enable a concise description of the slender body
theory in \cref{sec:hydro}. Defining $e=\sqrt{1-\epsilon^2}$ to be the
effective filament eccentricity, on the first segment we take
\begin{equation}
	\vec{f}(s) = \left\{\begin{array}{ll}
		\vec{f}_1\,, & s\in[s_1,s_L^{\star})\\
		\vec{f}_{2}\,, & s\in[s_L^{\star},s_{2})
	\end{array}\right. \quad\text{for } s_L^{\star} = \frac{1}{2}\left(\frac{L(1-e)}{2}+\ds\right)\,,
\end{equation}
whilst on the last segment we analogously have
\begin{equation}
	\vec{f}(s) = \left\{\begin{array}{ll}
		\vec{f}_N\,, & s\in[s_N,s_R^{\star})\\
		\vec{f}_{N+1}\,, & s\in[s_R^{\star},s_{N+1})
	\end{array}\right. \quad\text{for } s_R^{\star} = \frac{1}{2}\left(\frac{L(1+e)}{2}+\ds\right)\,.
\end{equation}
Whilst this is somewhat cumbersome, with the first and last segments being
treated differently to the others, we have found that it yields significant
advantages over simpler piecewise constant and linear schemes found in the
literature. In particular, attempts at a piecewise linear approximation, as in
\citet{Walker2019d}, result in large endpoint oscillations in the computed
values of $\vec{f}$, akin to those found in the regularised Stokeslet segment
methodology of \citet{Cortez2018} and are examined further in
\cref{sec:verify_examples:filament_in_flow}, where we evidence a lack of such
oscillations in the approach presented in this study. A natural alternative,
in which $\vec{f}$ is constant on each segment, yields equivalently
undesirable results, with the methodology becoming numerically intractable due
to stiffness when considering nearly straight filaments. Indeed, the same
issue is present in the scheme proposed by \citet{Hall-McNair2019}, which
utilises this intuitive discretisation. Though these issues are circumvented
by the approach presented in this work, the source of this sensitive numerical
dependence of the filament problem on discretisation remains unclear, and
warrants future investigation.

Returning to the now-discretised filament problem, the force-density
dependence of \cref{eq:integrated_balance_force,eq:integrated_balance_moment}
may be cast as a simple linear operator, denoted by $\tensor{B}$, allowing us to write
\begin{equation}\label{eq:BFR}
	-\tensor{B}\vec{F} = \vec{R}\,,
\end{equation}
where $\vec{R}=[0,0,m(s_1),\ldots,m(s_N)]^T$ encodes the bending moments and
total force acting on the filament, whilst
$\vec{F}=[\vec{f}_1\cdot\ex,\vec{f}_1\cdot\ey,\ldots,\vec{f}_{N+1}\cdot\ex,
\vec{f}_{N+1}\cdot\ey]^T$ is the vector of discretised force densities. The
first two rows $\tensor{B}_1,\tensor{B}_2$ of $\tensor{B}$ represent total
force balance over the filament, and are given explicitly by
\begin{align}
	\tensor{B}_1 &= \frac{\ds}{2}[1+d,0,2-d,0,2,0,\ldots,2,0,2-d,0,1+d,0]\,,\\
	\tensor{B}_2 &= \frac{\ds}{2}[0,1+d,0,2-d,0,2,0,\ldots,2,0,2-d,0,1+d]\,,
\end{align}
where $d=L(1-e)/(2\ds)$. The remaining rows $\tensor{B}_{i+2}$ encode the
integrated moment balance equations for $i=1,\ldots,N$, the expressions for
which are given in \cref{app:B}.

We now suppose that an invertible linear operator $\tensor{A}$ may be
constructed such that
\begin{equation}\label{eq:hydro}
	\dot{\vec{X}} = \tensor{A}\vec{F}(1 + O(\epsilon))\,,
\end{equation}
which we will find explicitly in \cref{sec:hydro}. Upon substitution of
\cref{eq:hydro} into \cref{eq:BFR}, also making use of \cref{eq:QTX}, we
obtain the leading-order coarse-grained linear system
\begin{equation}
	-\tensor{B}\tensor{A}^{-1}\tensor{Q}\dot{\vec{\theta}} = \vec{R}\,,
\end{equation}
where $\tensor{B}$ encodes the integrated equations of elasticity,
$\tensor{A}$ represents the hydrodynamic relation between velocity and force
density, $\tensor{Q}$ links the kinematic descriptions of the filament, and
$\vec{R}$ is the elastic response of the filament to bending.

Finally, we non-dimensionalise lengths by filament half-length $L/2$, forces with $4EI/L^2$, and time with some characteristic time scale $T$. This yields the dimensionless system
\begin{equation}
	-E_h \hat{\tensor{B}}\hat{\tensor{A}}^{-1}\hat{\tensor{Q}}\dot{\hat{\vec{\theta}}} = \hat{\vec{R}}\,, \quad E_h = \frac{\pi\mu L^4}{2EI\,T}\,,
\end{equation}
where the notation $\hat{\cdot}$ denotes dimensionless quantities, and we note
that the rescaled arclength parameter is $\hat{s} = 2s/L \in[0,2]$. The
elastohydrodynamic number $E_h$ here is analogous to that of
\citet{Walker2019d}, though differs by a factor of 16 due to differing choices
of lengthscale. We have the explicit relations
\begin{equation}\label{eq:nondimvars}
	\tensor{B} = \frac{L^2}{4} \hat{\tensor{B}}\,, \quad \tensor{A} = \frac{1}{8\pi\mu}\hat{\tensor{A}}\,, \quad \tensor{Q}\dot{\vec{\theta}} = \frac{L}{2T}\hat{\tensor{Q}}\dot{\hat{\vec{\theta}}}\,, \quad \vec{R} = \frac{2EI}{L}\hat{\vec{R}}\,,
\end{equation}
between dimensional and dimensionless quantities, having multiplied the force
balance equations by $\ds/2$ and absorbed the dimensional scalings of
$x_1,y_1$ in to $\tensor{Q}$ for covenience, writing $\hat{\vec{\theta}} =
(\hat{x}_1,\hat{y}_1,\theta_1,\ldots,\theta_N)^T$. In what follows, we will
drop the $\hat{\cdot}$ notation for dimensionless variables, though for later
convenience we first write
\begin{equation}
	\eta(s) = \frac{L}{2}\hat{\eta}(\hat{s}) =
	\frac{\epsilon L}{2}\tilde{\eta}(\hat{s})
\end{equation}
and immediately drop the tilde on $\tilde{\eta}(\hat{s})\sim O(1)$. For clarity, the
points on the surface of the filament may now be written in terms of
dimensionless quantities as
\begin{equation}
	\vec{x}^S(s,\phi) = \vec{x}(s) + \epsilon\eta(s)\vec{e}_r(s,\phi)\,.
\end{equation}

\subsection{Non-uniform hydrodynamics}\label{sec:hydro}
Before describing the slender-body theory of \citet{Walker2020b} that we will
use to relate forces and flow, we first recapitulate the well-known
regularised singularities of \citet{Cortez2001} and \citet{Ainley2008} on
which it is built. Following \citet{Walker2020b}, for points
$\vec{\alpha},\vec{\beta}$ and particular choices of mollifier the regularised
Stokeslet $\tensor{S}^{\chi}$ and potential dipole $\tensor{D}^{\chi}$ are
given by
\begin{eqnarray}
\tensor{S}^{\chi}(\vec{\alpha},\vec{\beta}) = \frac{(\abs{\vec{\alpha}-\vec{\beta}}^2 
	+ 2\chi)\tensor{I}}{(\abs{\vec{\alpha}-\vec{\beta}}^2 + \chi)^{3/2}} +
	  \frac{\tensor{Q}(\vec{\alpha},\vec{\beta})}{(\abs{\vec{\alpha}-\vec{\beta}}^2+\chi)^{3/2}}\,,
	  \label{regStokeslet} \\
	\tensor{D}^{\chi}(\vec{\alpha},\vec{\beta}) = -\frac{(\abs{\vec{\alpha}-\vec{\beta}}^2 - 2\chi)\tensor{I}}{(\abs{\vec{\alpha}-\vec{\beta}}^2
	+ \chi)^{5/2}} + \frac{3\tensor{Q}(\vec{\alpha},\vec{\beta})}{(\abs{\vec{\alpha}-\vec{\beta}}^2+\chi)^{5/2}}\,,
	\label{regdipole}
\end{eqnarray}
where $\tensor{Q}(\vec{\alpha},\vec{\beta}) =
(\vec{\alpha}-\vec{\beta})\otimes(\vec{\alpha}-\vec{\beta})$, $\tensor{I}$ is the $3\times3$
identity tensor, and $\chi$ is the regularisation parameter.

Throughout this section, it will be convenient to consider functions of
filament arclength instead as functions of a shifted arclength parameter
$s'\in[-1,1]$, which will greatly simplify the notation associated with the
slender-body theory of \citet{Walker2020b}. We will consistently abuse
notation and write $\vec{x}(s) \equiv \vec{x}(s')$, where $s=s'+1$ and other
functions of arclength are treated analogously. In particular, this enables us
to concisely define the arclength-dependent regularisation parameter
$\chi=\chi(s')$, which may be written as
\begin{equation}\label{eq:chi}
	\chi(s') = \epsilon^2[(1-s'^2) - \eta^2(s')]\,.
\end{equation}
Recalling the effective filament eccentricity as $e=\sqrt{1-\epsilon^2}$, the
dimensionless ansatz of \citet{Walker2020b} for the fluid velocity at a point
$\vec{y}$ in terms of the force per unit length $\vec{f}(s')$ may now be
written as
\begin{equation}\label{eq:ansatz}
	\vec{u}(\vec{y}) =
	\int\limits_{-e}^{e}\left[\tensor{S}^{\chi(s')}(\vec{y},\vec{x}(s'))
	-
	\frac{1-e^2}{2e^2}(e^2-s'^2)\tensor{D}^{\chi(s')}(\vec{y},\vec{x}(s'))\right]\vec{f}(s')\intd{s'}\,,
\end{equation}
noting that the dimensional factor of $8\pi\mu$ has been absorbed by the
scalings of \cref{eq:nondimvars}. Note that the limits in the integral are
between $-e$ and $e$, rather than $-1$ and $1$, as inherited ultimately from
the \citeauthor{Chwang1975} solution for a translating prolate ellipsoid, as
detailed in \citet{Walker2020b}. Taking $\vec{y}=\vec{x}^S(s_i',\phi)$, where
$s_i' = s_i - 1$ are the shifted dimensionless arclengths corresponding to the
discrete points $s_i$, we may apply this ansatz at the filament surface to
generate the $N+1$ vector equations
\begin{multline}\label{eq:sbt_surface}
	\vec{u}(\vec{x}^S(s_i',\phi)) =
	\int\limits_{-e}^{e}\left[\tensor{S}^{\chi(s')}(\vec{x}^S(s_i',\phi),\vec{x}(s'))\vphantom{\frac{1-e^2}{2e^2}}\right.\\\left.
	-
	\frac{1-e^2}{2e^2}(e^2-s'^2)\tensor{D}^{\chi(s')}(\vec{x}^S(s_i',\phi),\vec{x}(s'))\right]\vec{f}(s')\intd{s'}\,.
\end{multline}
We impose the no-slip condition $\vec{u}(\vec{x}^S(s_i',\phi)) =
\dot{\vec{x}}^S(s',\phi)$ on the surface of the filament, and may decompose
\begin{equation}
	\dot{\vec{x}}^S(s',\phi) = \dot{\vec{x}}(s') + \epsilon\vec{\omega}(s')\eta(s')\cross\vec{e}_r(s',\phi)\,, 
\end{equation}
for centreline velocity $\dot{\vec{x}}(s')$ and angular velocity
$\vec{\omega}(s')$ measured about $\vec{x}(s')$, recalling that the filament
is assumed to be unshearable. Supposing that $\vec{\omega}$ is $O(1)$ as
$\epsilon\rightarrow0$, consistent with the filament being assumed untwistable
and planar, at leading order we simply have
\begin{equation}
	\dot{\vec{x}}^S(s',\phi) = \dot{\vec{x}}(s') + O(\epsilon)\,,
\end{equation}
independent of $\phi$. Finally, we arrive at the leading-order relation
\begin{multline}\label{eq:leading_order_surface}
		\dot{\vec{x}}(s_i') \approx
	\int\limits_{-e}^{e}\left[\tensor{S}^{\chi(s')}(\vec{x}^S(s_i',\phi),\vec{x}(s'))\vphantom{\frac{1-e^2}{2e^2}}\right.\\\left.
	-
	\frac{1-e^2}{2e^2}(e^2-s'^2)\tensor{D}^{\chi(s')}(\vec{x}^S(s_i',\phi),\vec{x}(s'))\right]\vec{f}(s')\intd{s'}\,.
\end{multline}
For comparison, the equivalent expression used in the method of regularised
Stokeslet segments, and indeed many regularised slender body theories
\citep{Gillies2009,Cortez2012,Olson2013}, may be written as
\begin{equation}\label{eq:bad_ansatz}
		\dot{\vec{x}}(s_i') \approx \int\limits_{-1}^{1}
		\tensor{S}^{\frac{\epsilon}{2}}(\vec{x}(s_i'),\vec{x}(s'))\vec{f}(s')
		\intd{s'}\,,
\end{equation}
where we note in particular that the evaluations of the regularised Stokeslet
kernel are on the filament centreline, not on the surface of the slender body.

Though the left-hand side of \cref{eq:leading_order_surface} is trivially
independent of the cross-sectional angle $\phi$, it is not clear if the
integral is similarly independent. However, with the particular choice of
regularisation parameter $\chi(s')$ given in \cref{eq:chi},
\citet{Walker2020b} showed that the integral of \cref{eq:sbt_surface} is in
fact independent of $\phi$ at leading order in $\epsilon$, with errors linear
in the aspect ratio. Thus, \cref{eq:leading_order_surface} satisfies this
necessary condition. Moreover, its solution enables the no-slip condition on
the surface of the filament to be satisfied to $O(\epsilon)$. With the force
density $\vec{f}$ discretised as described in \cref{sec:coarse_graining} and
incurring errors proportional to $\ds\norm{\mathrm{d}\vec{f}/\mathrm{d}s}$,
which are assumed small and may be verified a posteriori, yields the
leading-order linear system
\begin{equation}
	\dot{\vec{X}} = \tensor{A}\vec{F}
\end{equation}
relating force densities on the filament to the centreline velocities.

% In order to write this relation explicity, we first define the effective
% segment endpoints as $\tilde{s}_1 = -e$, $\tilde{s}_{N+1} = e$, and
% $\tilde{s}_j = s_j'$ for $j=2,\ldots,N$, and note that these differ from the
% $s_j$ only for $j=1$ and $j=N$. From the $\tilde{s}_j$ we further define the
% effective midpoints $m_j = (\tilde{s}_j + \tilde{s}_{j+1})/2$ for
% $j=1,\ldots,N$, so that over the domain of integration, $s'\in[-e,e]$, we have
% \begin{equation}
% 	\vec{f}(s') = \left\{
% 	\begin{array}{lc}
% 	\vec{f}_1\,, & \tilde{s}_1 \leq x < m_1\,, \\
% 	\vec{f}_j\,, & m_{j-1} \leq x < m_j\,, \\ 
% 	\vec{f}_{N+1}\,, & m_N \leq x < \tilde{s}_{N+1}\,,
% 	\end{array}\right.
% \end{equation}
% recalling the analogous dimensional discretisation of
% \cref{sec:coarse_graining}. With these definitions, the discretised form of
% \cref{eq:leading_order_surface} may be written simply as
% \begin{equation}
% 	\dot{\vec{x}}(s_i') = \int\limits_{\tilde{s}_1}^{m_1}\tensor{g}\intd{s'} \vec{f}_1
% 		+ \sum\limits_{j=2}^{N} \int\limits_{m_{j-1}}^{\tilde{m}_j}\tensor{g}\intd{s'} \vec{f}_j 
% 		+ \int\limits_{m_N}^{\tilde{s}_{N+1}}\tensor{g}\intd{s'} \vec{f}_{N+1}\,,
% \end{equation}
% where $\tensor{g} = \tensor{g}(s',s_i',\phi)$ denotes the factor multiplying
% $\vec{f}$ in \cref{eq:leading_order_surface}. In writing the above we have
% also assumed that there are no discrete arclengths $s_j$ between $-1$ and $-e$
% or $e$ and $1$, though the approach may be readily modified to accommodate
% such discretisations.

\subsection{Regularised non-uniform segments}\label{sec:RNS}
The entries of $\tensor{A}$ may be readily computed with quadrature, as was
performed in the original work of \citet{Walker2020b}. However, with the
integral kernels rapidly varying in some regions, this can be prohibitively
expensive in elastohydrodynamic simulations, where numerous evaluations of
$\tensor{A}$ are typically required. Inspired by the recent method of
regularised Stokeslet segments, as detailed by \citet{Cortez2018} and used in
the work of \citet{Walker2019d}, we seek to evaluate these integrals
analytically on each linear segment of the filament, in general incurring
discretisation errors as a result of the arclength-dependent regularisation
$\chi(s')$. We will refer to this approach as the method of \emph{regularised
non-uniform segments} (RNS).

With $\vec{f}$ approximated as piecewise constant, computation of $\tensor{A}$
reduces to performing a number of integrals over the discrete straight
segments of the filament. We consider such an integral over the
$j$\textsuperscript{th} segment, parameterised by $\alpha\in[0,1]$, on which
we write $\chi=\chi(\alpha)$ and the filament centreline is given by
$\vec{x}(\alpha)=\vec{x}_j -
\alpha\vec{v}$, where $\vec{v}=\vec{x}_j-\vec{x}_{j+1}$. This formulation is
applicable even to the first and last segments, subject to the substitution of
$\vec{x}_1$ by $\vec{x}(-e)$ and of $\vec{x}_{N+1}$ by $\vec{x}(e)$, owing to
the chosen discretisation of $\vec{f}$. With this discretised $\vec{f}$ taking
the values $\vec{f}_j$ and $\vec{f}_{j+1}$ on different halves of this
segment, we require expressions for the integrals evaluated on two subdomains,
$\alpha\in[0,1/2]$ and $\alpha\in[1/2,1]$. Further, analogous to
\citet{Cortez2018} and \citet{Walker2019d} and given explicitly in
\cref{app:linear_combination}, we note that this discretisation of $\vec{f}$
has rendered each of these integrals as linear combinations of
\begin{equation}
	T_{m,q}^L = \int\limits_{0}^{\frac{1}{2}}\alpha^mR^q\intd{\alpha}\,, \quad T_{m,q}^R = \int\limits_{\frac{1}{2}}^1\alpha^mR^q\intd{\alpha}\,,
\end{equation}
where we define $R=\left(\abs{\vec{x}^S_i-\vec{x}_j+\alpha\vec{v}}^2 +
\chi(\alpha)\right)^{1/2}$ for a surface point
$\vec{x}^S_i=\vec{x}^S(s_i',\phi)$, with $\phi$ arbitrary as above. These
integrals may be readily performed in the case that $R^2$ is a quadratic
function of $\alpha$, as in the original method of regularised Stokeslet
segments though here prohibited in general by $\chi(\alpha)$.

In order to recover this desirable property, we Taylor expand $\chi(\alpha)$
about an endpoint of the segment, either $\alpha=0$ or $\alpha=1$, assuming
sufficient smoothness of $\chi$. The expansion point is chosen in order to
minimise the error in the resulting integral, noting in particular that
$R(\alpha)$ can become $O(\epsilon)$ if $\vec{x}_j-\alpha\vec{v}$ nears
$\vec{x}_i^S$, for example in the trivial case of $i=j$. With $\chi$ therefore
plausibly the dominant term in this $O(\epsilon)$ neighbourhood, we choose to
expand $\chi(\alpha)$ about the segment endpoint that is closest to
$\vec{x}_i^S$, denoting the value of $\alpha$ at this endpoint as
$\alpha^{\star}$. Collecting powers of $\alpha$, in each case this yields an
expansion of the form
\begin{equation}\label{eq:R_expansion}
	R(\alpha)^2 = A + B \alpha + C\alpha^2 + E\,, \quad \abs{E} \leq
	\left\{\begin{array}{ll}
	\frac{1}{6}\alpha^3\abs{\vec{v}}^3\sup\abs{\diff{^3\chi}{s'^3}}\,, & \alpha^{\star}=0\,,\\
	\frac{1}{6}(1-\alpha)^3\abs{\vec{v}}^3\sup\abs{\diff{^3\chi}{s'^3}}\,, & \alpha^{\star}=1\,.
	\end{array}\right.
\end{equation}
In the error term $E$ we have bounded the third derivative of $\chi$ over the
segment, and have cast the derivative in terms of the normalised arclength
$s'$ in order to unify our phrasing of model assumptions. With
$R(\alpha)=O(\epsilon)$ when $\abs{\vec{x}_i^S
-\vec{x}_j+\alpha\vec{v}}=O(\epsilon)$, and $R(\alpha)$ strictly order unity
otherwise, when $\alpha^{\star}=0$ this error term is subdominant if
\begin{equation}\label{eq:chi_restrictions}
	\frac{1}{6}\alpha^3\ds^3\sup\abs{\diff{^3\chi}{s'^3}} = \left\{\begin{array}{ll}
	O(\epsilon^3)\,, & \text{where } \abs{\vec{x}_i^S-\vec{x}_j+\alpha\vec{v}}=O(\epsilon)\,,\\
	O(\epsilon)\,, & \text{otherwise},\end{array}\right.
\end{equation}
noting that $\abs{\vec{v}}\leq\ds$, with a similar expression required for
$\alpha^{\star}=1$. This imposes a weak restriction on the derivatives of
$\chi$ and the discretisation length $\ds$, recalling that
$\chi=O(\epsilon^2)$ everywhere. Assuming that such a restriction holds, we
drop the error term $E$ in what follows, approximating $R(\alpha)$ as a
quadratic function on each segment. The segment-dependent coefficients $A,B,C$
may be readily computed when expanding with $\alpha^{\star}=0$ or
$\alpha^{\star}=1$, and for $\alpha^{\star}=0$ are given explicitly by
\begin{equation}\label{eq:ABC}
	A = \abs{\vec{x}_i^S - \vec{x}_j}^2 + \chi \,, \quad
	B = 2\vec{v}\cdot(\vec{x}_i^S - \vec{x}_j) + \abs{\vec{v}}\diff{\chi}{s'}\,, \quad
	C = \abs{\vec{v}}^2\left(1 + \diff{^2\chi}{s'^2}\right)\,,
\end{equation}
where evaluations of $\chi$ and its derivatives for $A,B,C$ are at $\alpha=0$
and we henceforth write $R(\alpha)^2 = A + B\alpha + C\alpha^2$ for brevity.
Omitted here for brevity, analogous expressions hold for $A,B,C$ when
$\alpha^{\star}=1$. As noted above and written explicitly in
\cref{app:linear_combination}, the integral kernel may be decomposed into a
linear combination of terms $\alpha^mR^q$ for
$(m,q)\in\{(0,-1),(0,-3),(0,-5),(1,-3),(1,-5),(2,-3),(2,-5),(3,-5),(4,-5)\}$.
For $m>0$, computation of these quantities may be performed simply via the
recurrence relations
\begin{align}
	T_{m+1,q-2}^L &= \left.\frac{\alpha^mR^q}{qC}\right\rvert_0^{\frac{1}{2}} -\frac{m}{qC}T_{m-1,q}^L - \frac{B}{2C}T_{m,q-2}^L\,,\label{eq:recurrenceL}\\
	T_{m+1,q-2}^R &= \left.\frac{\alpha^mR^q}{qC}\right\rvert_{\frac{1}{2}}^1 -\frac{m}{qC}T_{m-1,q}^R - \frac{B}{2C}T_{m,q-2}^R\,,\label{eq:recurrenceR}
\end{align}
where $q,C\neq0$. These are analogous to the recurrence of \citet{Cortez2018}
and are similarly derived via integration by parts. Thus, explicit calculation
of $T_{m,q}^L$ and $T_{m,q}^R$ is required only for $m=0$, with the relevant antiderivatives given in \cref{app:antiderivatives}.

Hence, the construction of the operator $\tensor{A}$ proceeds simply and
efficiently: the coefficients $A,B,C$ are evaluated from precomputed values of
$\chi$ and its derivatives, the integrals $T_{m,q}^L,T_{m,q}^R$ are computed
for $m=0$ using the given antiderivatives, further integrals for $m>0$ are
computed via the recurrences of \cref{eq:recurrenceL,eq:recurrenceR}, and the
entries of $\tensor{A}$ are formed as linear combinations of these terms
following \cref{app:linear_combination}. We additionally note that this
process may be readily generalised to evaluation points that do not lie on the
surface of the filament, in this case Taylor expanding about the segment
endpoint that is closest to the evaluation point.
\section{Verification and Examples}
\label{sec:verify_examples}
%autoignore
\subsection{Efficiency and accuracy against quadrature}
\begin{figure}
	\centering
	\vspace{1em}
	\begin{tabular}{lccccc}
		&\includegraphics[width=0.25\textwidth]{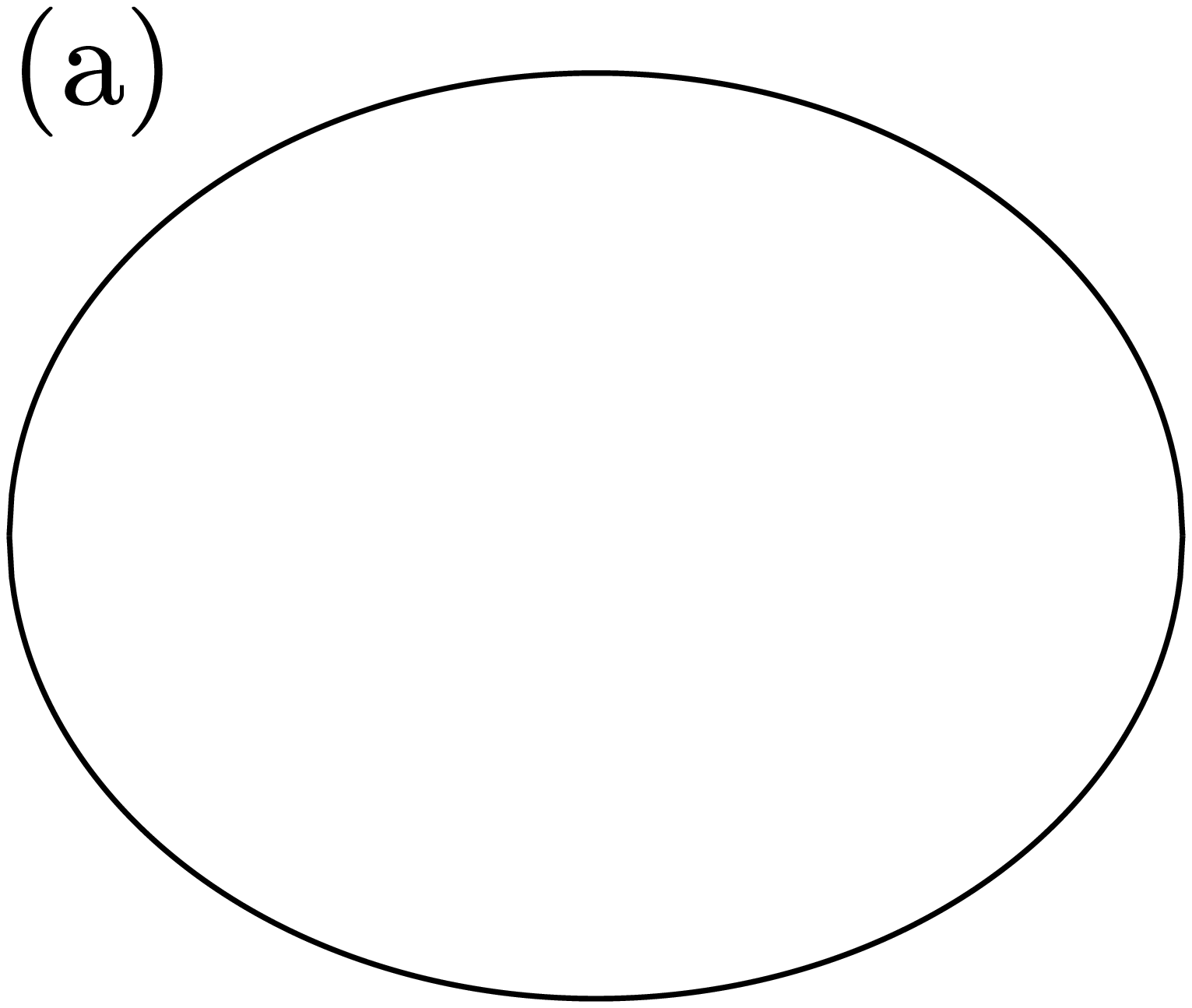} & \includegraphics[width=0.25\textwidth]{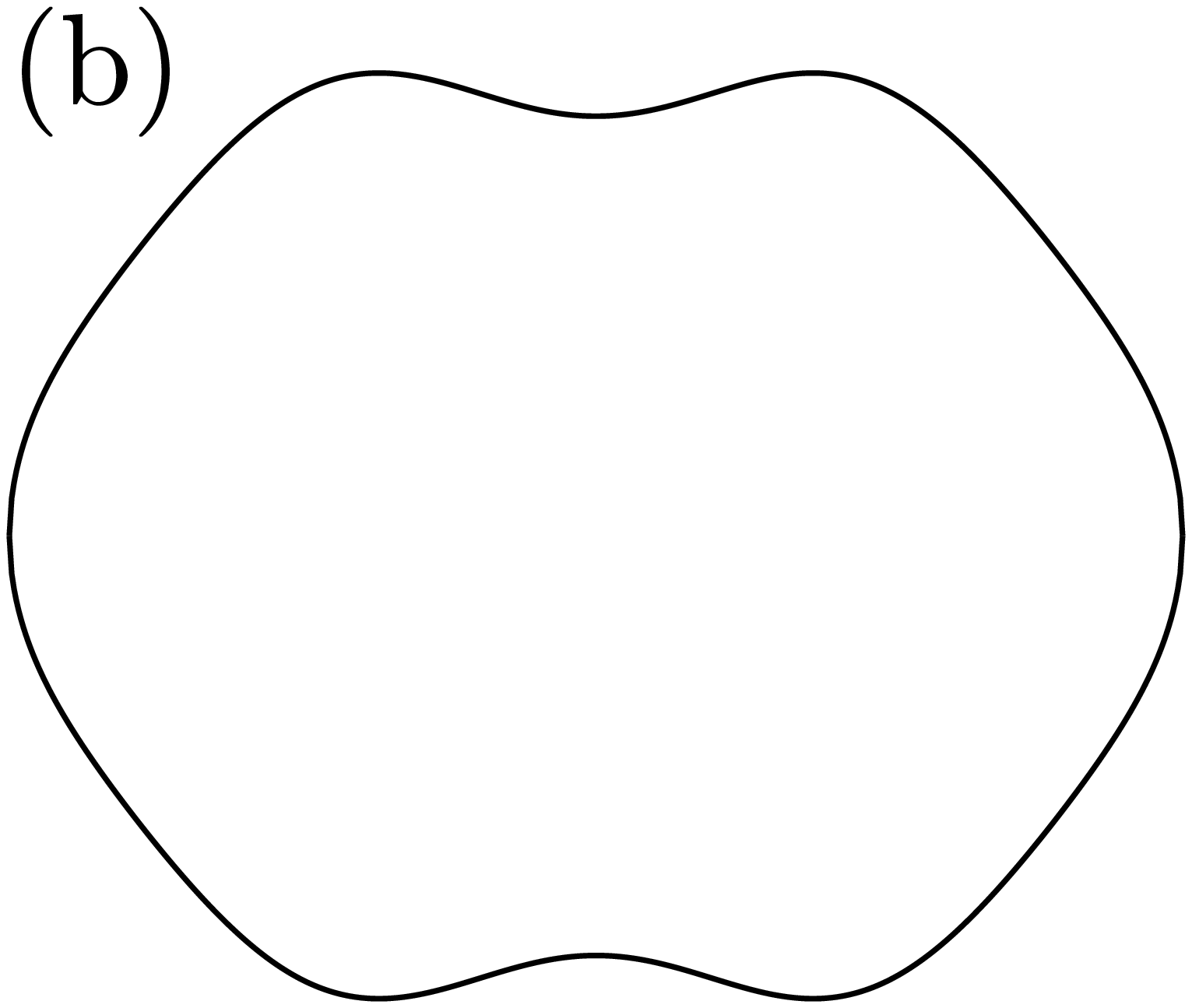} & \includegraphics[width=0.25\textwidth]{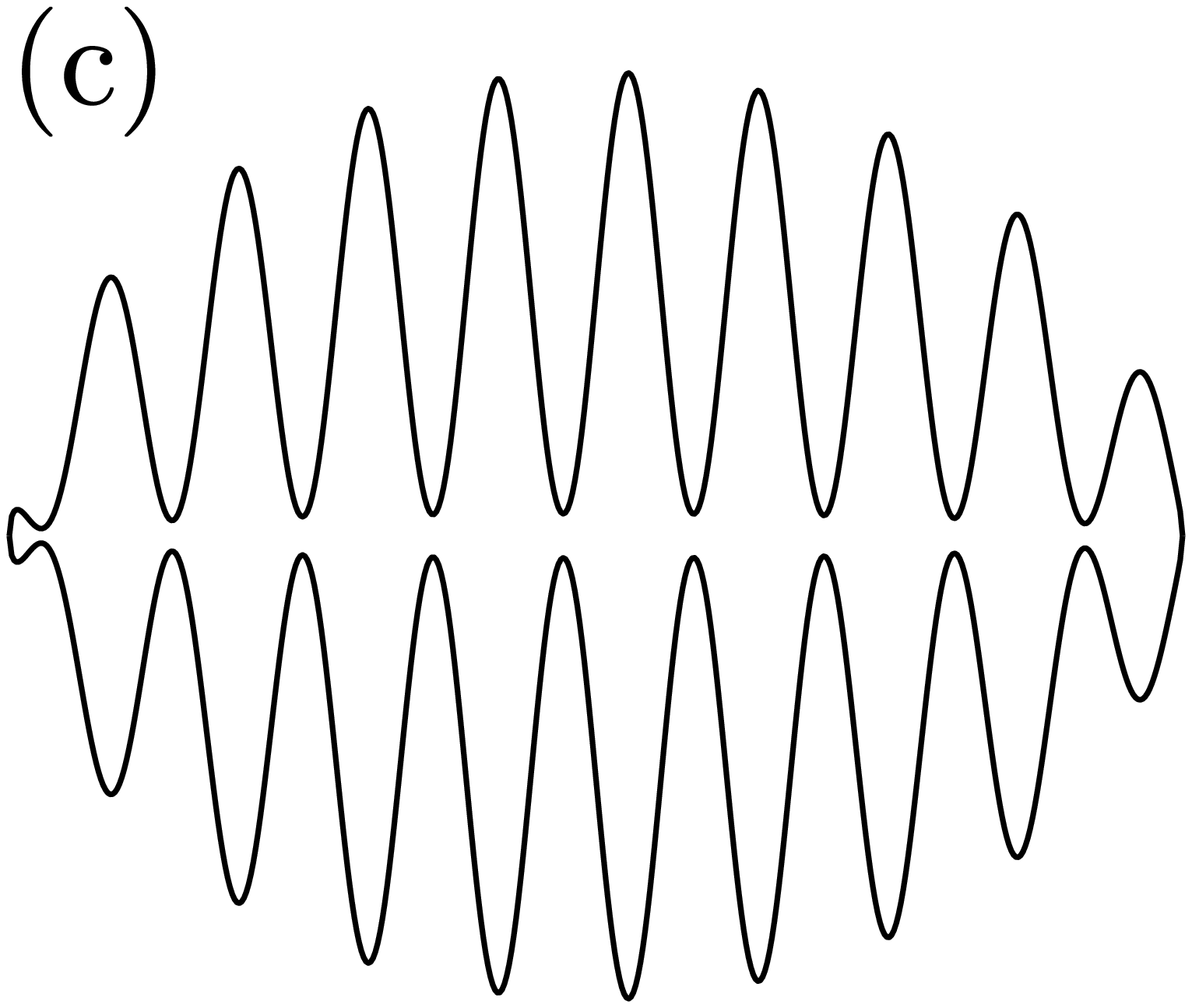}\\
		$\mathcal{E}$ & \SI{3.2e-11} & \SI{3.3e-6}{} & \SI{5.4e-4}{}\vspace{0.3em}
	\end{tabular}
	\caption{Example radius functions, and the relative error $\mathcal{E}$ of
	using the RNS method in constructing the operator $\tensor{A}$ compared
	with a quadrature rule of tolerance \SI{e-12}{}. In each of the three
	cases we note a small matrix infinity norm error $\mathcal{E}$, largest in
	case (c) where curvature of the radius function is rapidly varying. Here
	we have considered a curved filament in a dimensionless framework with
	$N=100$ segments, having taken $\epsilon=0.02$ and radius functions
	corresponding to \cref{eq:eta}. The filament centreline corresponds to the
	initial condition of \cref{fig:relax}a, and shapes are shown stretched
	vertically for visual clarity.}
	\label{fig:eta}
\end{figure}

Construction of the operator $\tensor{A}$ via the method of regularised
non-uniform segments introduces local approximations of the regularisation
parameter $\chi$ wherever it is not simply a quadratic function of arclength,
enabling analytic integration. We now compare this approach with quadrature in
terms of both accuracy and efficiency in a practical parameter regime,
considering three dimensionless radius functions $\eta(s')$ of varying
complexity:
\begin{equation}\label{eq:eta}\begin{array}{ll}
	\text{(a)} & \sqrt{1-s'^2}\,,\\
	\text{(b)} & \sqrt{1-s'^2}(1-0.1\cos{2\pi s'})\,,\\
	\text{(c)} & \sqrt{1-s'^2}(1.1+\sin{9\pi s'})\,,
	\end{array}
\end{equation}
each subject to normalisation and shown in \cref{fig:eta}. Considering a
filament with a curved centreline, corresponding to the initial condition of
\cref{fig:relax}a, with $N=100$ and $\epsilon=0.02$ we compute $\tensor{A}$
using both the RNS methodology and the inbuilt \texttt{quadv} routine in
MATLAB{\textsuperscript\textregistered}, with the numerical quadrature set to
a tolerance of \SI{e-12}{} and denoting the results of these computations by
$\tensor{A}_{\text{RNS}}$ and $\tensor{A}_{\text{Q}}$, respectively. We write
$\mathcal{E}$ for the relative matrix infinity norm error between these two
results, defined explicitly as
\begin{equation}
	\mathcal{E} = \frac{\norm{\tensor{A}_{\text{RNS}} - \tensor{A}_{\text{Q}}}_{\infty}}{\norm{\tensor{A}_{\text{Q}}}_{\infty}}\,.
\end{equation}
These relative errors are shown in \cref{fig:eta}, each of which can be seen
to be several orders of magnitude lower than the asymptotic slenderness
parameter. The rapidly varying curvature of case (c) gives rise to the largest
error, consistent with the restrictions imposed on the derivatives of $\chi$
in \cref{eq:chi_restrictions}. Computations were performed on modest hardware
(Intel\textsuperscript{\textregistered} Core\texttrademark\ i7-6920HQ CPU),
with the walltime for the RNS method being over two orders of magnitude less
than that of the quadrature implementation, representing a significant
improvement in computational efficiency for minimal reduction in accuracy.
These observations of efficiency and accuracy hold for a range of considered
body centrelines and radius functions, and are robust to variations in the
slenderness parameter $\epsilon$.

\subsection{Invariants of free-filament motion}
\begin{figure}
	\centering
	\vspace{1em}
	\begin{overpic}[percent,width=\textwidth]{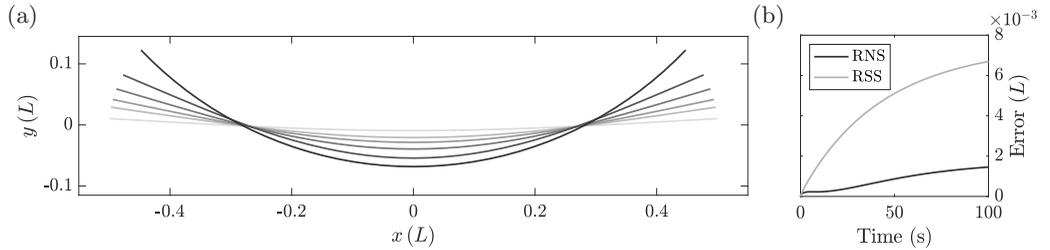}
	\put(-1,22){(a)}
	\put(72,22){(b)}
	\end{overpic}
	\caption{The relaxation of a symmetric filament, simulated with $N=40$
	segments for $E_h=9600$. (a) Relaxation dynamics qualitatively match those
	of \citet{Walker2019d}, in agreement with intuition and preserving the
	symmetry of the initial condition. (b) Distance translated by the centre
	of mass of the filament, as computed by the presented RNS methodology and
	the RSS approach of \citet{Walker2019d}, analytically zero and captured
	approximately here, having taken $N=40$ and $\epsilon=0.02$. Here we have
	considered a filament with dimensionless shape $\eta(s')=\sqrt{1-s'^2}$,
	corresponding to a prolate ellipsoid, though note that this information is
	not captured by the typical slender body ansatz, as implemented in
	\citet{Walker2019d}.}
	\label{fig:relax}
\end{figure}

The coarse-grained framework for filament elasticity is similar to that
presented and derived in the recent work of \citet{Walker2019d}, where it was
extensively verified and benchmarked, utilising the stiff solver
\texttt{ode15s} provided in MATLAB{\textsuperscript\textregistered} with
relative and absolute tolerances of \SI{e-6}{} \citep{Shampine1997}. However,
due to the modification of considering a piecewise constant discretisation of
the force density $\vec{f}$, akin to the study of \citet{Moreau2018}, we
additionally verify the presented methodology in the case of a relaxing
symmetric filament. Having taken $N=40$ and $\epsilon=0.01$, in
\cref{fig:relax} we showcase the simulated dynamics of an initially symmetric
filament relaxing to a straight configuration, during which we see that
symmetry is preserved. Owing to the filament having no net force or torque act
upon it, the centre of mass should not deviate from its initial position.
Computing the translation of the centre of mass over the motion, a
quantitative measure of framework accuracy, in \cref{fig:relax}b we see that
this approximate constancy is preserved numerically with errors on the order
of $\SI{e-3}{}L$, improved by an order of magnitude when compared to the
previous methodology of \citet{Walker2019d}.

\subsection{Comparison against existing theories}
We now more thoroughly compare and contrast the presented elastohydrodynamic
framework against two existing approaches, in particular the published RSS
methodology of \citet{Walker2019d} and a resistive force theory (RFT)
formulation based on that of \citet{Moreau2018}. The latter RFT method is as
described in the work of \citet{Walker2019d}, though we make use of the
resistive coefficients of \citet{Hancock1953,Gray1955}, with the normal
resistive coefficient twice that of the tangential coefficient.

\subsubsection{A relaxing filament}
\begin{figure}
	\centering
	\vspace{1em}
	\begin{overpic}[percent,width=\textwidth]{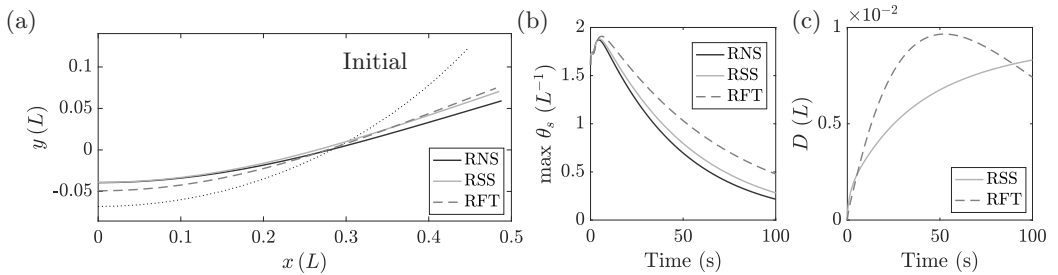}
	\put(-2,24){(a)}
	\put(48,24){(b)}
	\put(75,24){(c)}
	\put(31,20){\footnotesize{Initial}}
	\end{overpic}
	\caption{Comparing methodologies via the relaxation of a symmetric filament, simulated with $N=40$
	segments for $E_h=9600$. Each starting from an initial curved
	configuration, shown dotted in (a), we simulate the relaxation dynamics
	via a resistive force theory (RFT), regularised Stokeslet segment (RSS),
	and regularised non-uniform segment (RNS) methodology. (a) Shown at the
	same instant in time (\SI{30}{\second}) are the filament configurations as
	computed by the three methodologies, with the filament shapes broadly
	similar though showing some minor differences. Only half of the filament
	is shown, appealing to the preserved symmetry, and the shared initial
	condition is shown as a dotted curve. (b) The maximum filament curvature
	as a function of time, highlighting greater distinctions between the
	methodologies. (c) The difference between filament configurations at time
	$t$, defined by $D^2 =
	\int\abs{\vec{x}_{\text{O}} -
	\vec{x}_{\text{RNS}}}_2^2\intd{s}/L$, quantifies the difference between
	the RNS method, denoted $\vec{x}_{\text{RNS}}$, and the results of the
	other frameworks, denoted $\vec{x}_{\text{O}}$. With a filament aspect
	ratio of 1:100 here, overall differences between computations appear only
	slight, with the exception of the longer timescale of the RFT solution
	compared to the non-local methodologies. In the RNS framework we have
	considered a filament with dimensionless shape $\eta(s')=\sqrt{1-s'^2}$,
	corresponding to a prolate ellipsoid, though note that this information is
	not captured by the RFT or RSS frameworks.}
	\label{fig:relax_all}
\end{figure}

We simulate the free relaxation of a bent filament, with the $\theta_i$
initially equally spaced and increasing between $-\pi/4$ and $\pi/4$ to
correspond to a filament of constant curvature, via each of the three
methodologies, picking a common but arbitrary elastohydrodynamic number of
$E_h=9600$ and setting $N=40$. The filament has aspect ratio 1:100,
corresponding to $\epsilon=0.02$ in the RNS framework and $\epsilon=0.01$ in
the RFT and RSS approaches. Simulating until a dimensional time of
\SI{100}{\second}, at which point the RNS solution is nearing complete
relaxation to a straight configuration, we display snapshots of the computed
solutions and some associated metrics in \cref{fig:relax_all}. Immediately
evident is a qualitative similarity between the computations, though there is
some pairwise disagreement throughout the motion. Most prominent are
differences in the timescale of relaxation, as can be seen in the maximum
curvature plot of \cref{fig:relax_all}b, with the RFT solution relaxing more
slowly than the predictions by non-local theories. We more concretely quantify
the overall differences between methodologies at a given time $t$ via the
measure $D$, defined for a computed solution $\vec{x}(s,t)$ by
\begin{equation}
	D^2(t) = \frac{1}{L}\int\limits_0^L
	\abs{\vec{x}(s,t) -
	\vec{x}_{\text{RNS}}(s,t)}_2^2\intd{s}\,,
\end{equation}
relative to the RNS solution $\vec{x}_{\text{RNS}}$. The evolution of this
distance measure for the RFT and RSS approaches is shown in
\cref{fig:relax_all}c, and demonstrates that, whilst differences between
solutions are indeed small, being on the scale of $\epsilon$ in this
particular case, these distinctions persist throughout the motion.

With elastohydrodynamic simulations appearing broadly similar at the level of
detail considered thus far, we also note a common computational efficiency of
the frameworks, with even the more complex regularised non-uniform segments
approach computing the relaxation dynamics in a number of seconds. Indeed,
this is replicated throughout further testing for each of a wide array of
initial conditions, and is robust to variations in the radius function
$\eta(s')$ and the filament aspect ratio. Thus, despite employing a more
sophisticated slender-body ansatz, we see retained in the RNS methodology the
desirable efficiency associated with the existing coarse-grained frameworks.

\subsubsection{A simple filament in flow}\label{sec:verify_examples:filament_in_flow}
From the agreement seen above in the case of a relaxing filament, one might
expect that the theoretical refinement offered by the RNS approach over the
simpler and cruder RSS methodology is minimal in practice. However, more
significant differences are indeed present, as we now highlight via a simple
example.

We consider perhaps the most simple possible filament simulation: the dynamics
of an initially straight filament in a uniform background flow, with a
background flow $\vec{u}_b$ incorporated into the current framework via the
mapping $\vec{u}\mapsto\vec{u}-\vec{u}_{b}$ as in the work of
\citet{Walker2019d}. The simulated filament should exhibit trivial motion and
deformation, merely translating with the background flow and retaining its
straight configuration. Both the RNS and RSS methodologies successfully
replicate this behaviour, and solution time is negligible. However, a noted
issue of methods based on regularised Stokeslet segments and similar
approaches are endpoint oscillations in the computed force density $\vec{f}$,
present in each of the works of
\citet{Cortez2018,Walker2019d,Hall-McNair2019}, which persist even with mesh
refinement.

\begin{figure}
	\centering
	\vspace{1em}
	\begin{overpic}[percent,width=\textwidth]{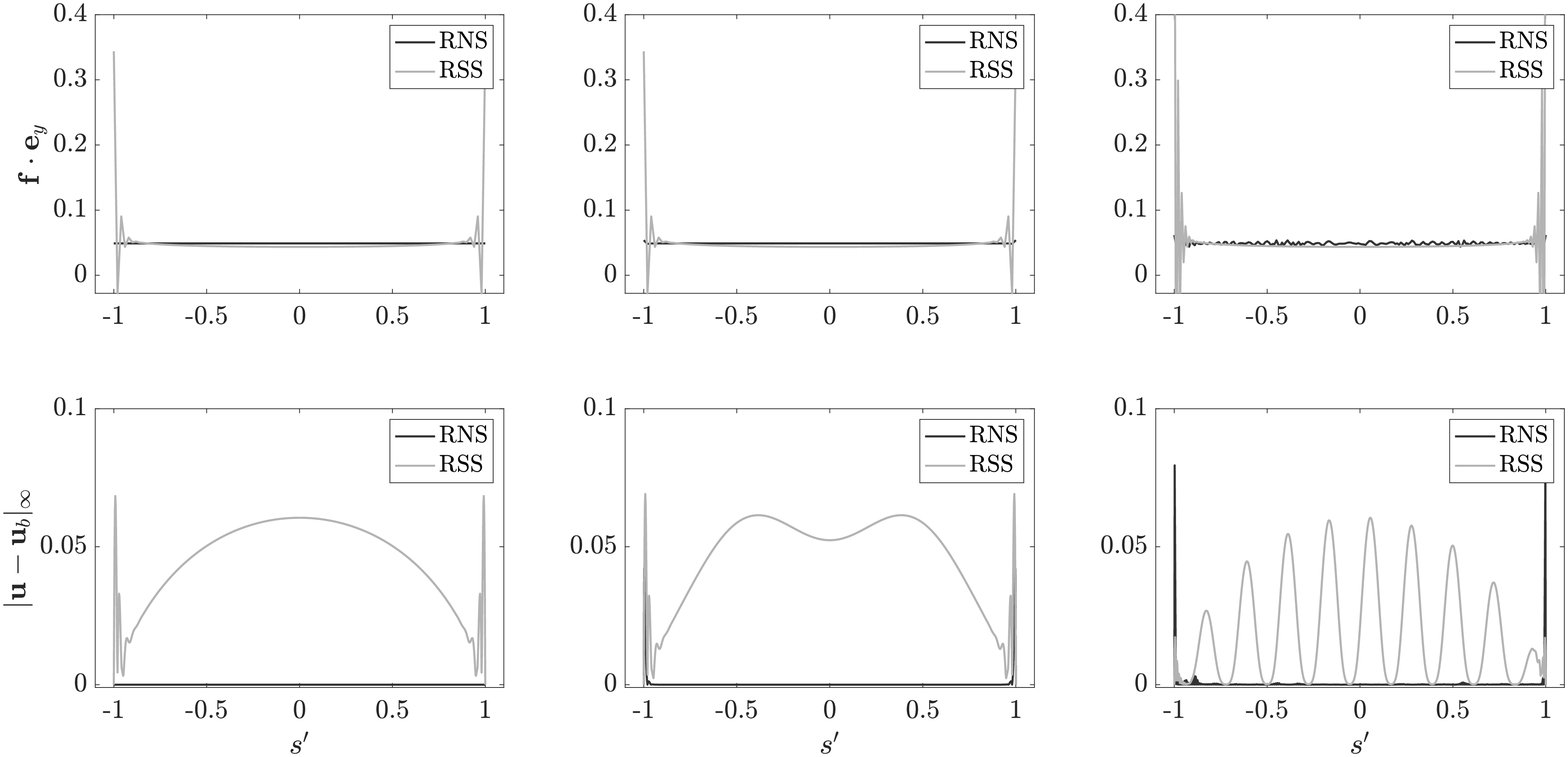}
	\put(0,49){(a)}
	\put(34,49){(b)}
	\put(68,49){(c)}
	\put(0,24){(d)}
	\put(34,24){(e)}
	\put(68,24){(f)}
	\end{overpic}
	\caption{The computed force densities and errors in surface velocity for
	straight filaments in unit uniform flow, with shapes corresponding to
	\cref{fig:eta}. Here we have used an aspect ratio of 1:100, corresponding
	to $\epsilon=0.02$ for the RNS methodology, and we recall that the slender
	body theory upon which it is based is accurate to $O(\epsilon)$. In panels
	(a-c) we note the presence of significant oscillations near the ends of
	the filament for the RSS solution, absent from the RNS computation. Panels
	(d-f) report the error in the surface velocity for a unit magnitude
	background flow $\vec{u}_b=\vec{e}_y$, from which we note the significant
	improvement in accuracy afforded by the RNS methodology over the RSS
	approach. In particular, the RNS error is at least an order of magnitude
	less than the RSS error, except perhaps at the very endpoints of the
	filament, with the RSS methodology making little systematic attempt to
	satisfy the boundary condition on the surface. We have taken $N=100$ in
	(a,b,d,e), whilst in (c) and (f) we have taken $N=200$, with the highly
	curved radius function of \cref{fig:eta}c requiring reduced $\ds$ to yield
	comparable accuracy to the other, simpler cases. Panels (a,d), (b,e),
	(c,f) correspond to the shapes shown in panels (a), (b), and (c) of
	\cref{fig:eta}.}
	\label{fig:force_no-slip}
\end{figure}

Here, we explicitly compute the force density on a straight filament of aspect
ratio 1:100 in a unit background flow $\vec{u}_b=\vec{e}_y$ using both the RNS
and RSS approaches, where $\vec{e}_y$ is perpendicular to the filament
tangent. In \cref{fig:force_no-slip}a-c we present the magnitude of the
computed force density on the filament from $s=0$ to $s=L$ for various body
radius functions, appealing to symmetry and noting that the force density is
identically zero in the direction of the filament tangent. In each case, we
observe the oscillations of the RSS force density near the endpoints of the
slender body, with the RSS solution being fundamentally independent of the
radius function, whilst the piecewise-constant RNS solution essentially
eliminates these oscillations. We have taken $N=200$ in
\cref{fig:force_no-slip}c in order to capture the highly oscillatory radius
function of \cref{fig:eta}c, consistent with the error analysis of
\cref{sec:RNS}, taking $N=100$ in the other cases.

Perhaps more pertinent, and indeed the motivation behind the use of the ansatz
of \citet{Walker2020b}, is the velocity boundary condition on the filament. We
explicitly evaluate the flow velocity on the surface of the filament via both
the RNS and RSS methods, sampling at 1000 uniformly spaced points on the
surface, and show the infinity norm error in the computed velocity in
\cref{fig:force_no-slip}d-f as a function of dimensionless shifted arclength
$s'$. Notably, the RSS approach is consistently inaccurate along the length of
the slender body, yielding approximately 5\% errors over the entire surface,
corresponding to five times the regularisation parameter of the RSS method.
The RNS methodology significantly improves upon this, with limitingly small
error along the majority of each of the slender bodies in both
\cref{fig:force_no-slip}d and \cref{fig:force_no-slip}e, with errors of
approximately $2\epsilon$ near the endpoints of the slender body in
\cref{fig:force_no-slip}e. In particular, these errors are on the same order
as those found in the original evaluation of the slender body theory by
\citet{Walker2020b}, with the impact of moving away from quadrature therefore
minimal in all but \cref{fig:force_no-slip}f, which is improved by reducing
$\ds$ to once again accommodate the oscillatory radius function. Thus, we
observe that the use of the RNS methodology affords significant gains in the
accuracy of the no-slip boundary condition over other approaches. Convergence
of this velocity error as a function of $N$ and $\epsilon$ is illustrated in
\cref{app:velocity_convergence} for the case of \cref{fig:eta}b.
\section{Discussion}
\label{sec:discussion}
%auto-ignore
Though the study of \citet{Moreau2018} vastly increased the computational
efficiency of filament simulations, it did so whilst employing only resistive
force theory, with this leading order hydrodynamic relation typically
conferring errors logarithmic in the filament aspect ratio. Subsequent works
have extended this framework to feature improved hydrodynamics
\citep{Walker2019d,Hall-McNair2019}, each making use of a simple but non-local
regularised ansatz. However, even these works neglect the boundary condition
on the body surface, instead evaluating velocities along the centreline when
linking fluid velocity to applied force density. Via the evaluations performed
in \cref{sec:verify_examples:filament_in_flow} of this work, we have evidenced
the relative inaccuracy of such approaches, observing non-negligible errors in
the computed surface velocity over the entire length of the slender body, with
these hydrodynamic errors being a fundamental weakness of previous
methodologies. Incorporating a refined hydrodynamic ansatz, the presented
regularised non-uniform segment methodology significantly improves upon such
errors, with discrepancies in the velocity boundary condition present only at
the filament endpoints, given adequate discretisation to account for the level
of variation in the cross-sectional radius function. In particular, the
slender-body theory employed here inherently takes into account the complex
shape of the filament, enabling the study of realistic slender-body geometries
and replacing previous imprecise justifications with analytically derived
quantifications of accuracy.

However, a naive incorporation of the slender-body theory of
\citet{Walker2020b} into a coarse-grained framework of filament elasticity
yielded large computation times, sacrificing the efficiency typically
associated with the underlying approach of \citet{Moreau2018}. Indeed, whilst
the use of automated quadrature rules allows computation of the hydrodynamic
operator to any desired degree of numerical accuracy, even the regular
integral kernel of the ansatz of \citet{Walker2020b} was insufficient to
enable rapid computation on par with the existing frameworks of
\citet{Moreau2018,Hall-McNair2019,Walker2019d}. Thus, exploiting a low-degree
approximation of the unknown force density $\vec{f}$, we instead computed the
necessary integrals analytically, mimicking the approach of \citet{Cortez2018}
after Taylor expanding the generally non-quadratic regularisation parameter
$\chi$. Quantifying the errors associated with this approximation, we have
evidenced a remarkable accuracy and efficiency of this approach, yielding a
scheme for elastohydrodynamic simulation that is comparable in computational
cost to existing methodologies, whilst simultaneously improving on their
accuracy. Thus, the presented framework will enable rapid solution of the
forward elastohydrodynamic problem, pertinent to modern Bayesian parameter
inference techniques, for example, along with explorations of fluid-structure
interactions in slender-body systems. Further, the method of regularised
non-uniform segments will more generally enable rapid application of the
slender-body theory of \citet{Walker2020b}, facilitating future investigative
and explorative studies into filament dynamics.

Whilst efficiency gains were made by adopting the general principle of the
method of regularised Stokeslet segments, the regularised non-uniform segment
approach avoids a pertinent issue associated with the principles of the former
theory. Present in the works and published codes of
\citet{Cortez2018,Walker2019d,Hall-McNair2019} are severe variations in the
computed force density $\vec{f}$ near the endpoints of the considered
filaments, persisting or indeed worsening with increased refinement of
approximation. With force density a fundamental component of such
elastohydrodynamic frameworks, these apparent errors may contribute
non-negligibly to simulated dynamics and applications, particularly given the
reported significance of distal activity in recent model spermatozoa
\citep{Neal2020}. Thus, the absence of comparable oscillations in the RNS
solutions represents a significant advantage over these existing
methodologies. Curiously, the insertion of the slender body theory of
\citet{Walker2020b} alone into the framework of \citet{Walker2019d} was not
sufficient to achieve this, as discovered during the author's initial attempt
at formulating the RNS methodology, which differs to the presented approach
only by using a piecewise linear discretisation of force density $\vec{f}$.
However, the combination of this improved ansatz and a lower order
discretisation of $\vec{f}$ successfully removed the unphysical oscillations
from the computed solutions, yielding the smooth profiles seen in
\cref{fig:force_no-slip}, though detailed investigation of the Fredholm
integral equation of \cref{eq:general_ansatz} is required in order to
ascertain the source of such pervasive errors. Future work may also include
trivial extensions to the study of active filaments and general background
flows, affording justified accuracy to the wide range of elastohydrodynamic
problems made tractable by the work of \citet{Moreau2018}.

In summary, we have integrated the fundamental advance of \citet{Moreau2018}
and the regularised slender-body theory of \citet{Walker2020b}, overcoming
their respective shortfalls to yield a framework for the efficient and
accurate simulation of slender-body elastohydrodynamics. The so-called
regularised non-uniform segment approach retains the flexibility of its parent
models, and hence may be applied to a wide variety of biological and
biophysical problems to afford increased accuracy over earlier approaches.
Further, complex axisymmetric geometries may now be reliably modelled using
this framework, previously only realisable with reduced fidelity or
drastically increased computational effort. Applicable even more generally,
this study has markedly improved the efficiency of the slender-body theory of
\citet{Walker2020b}, with this work overall facilitating both the accurate
quantification and large scale no-slip simulation of slender elasticity and
hydrodynamics.

% Acknowledgements.
B.J.W.\ is supported by the UK Engineering and Physical Sciences Research
Council (EPSRC), grant EP/N509711/1.

Declaration of Interests. The authors report no conflict of interest.

\appendix
%auto-ignore
\section{Moment balance as a linear system}
\label{app:B}
For $i=1,\ldots,N$, the rows $\tensor{B}_{i+2}$ of $\tensor{B}$ encode the
integrated moment balance in terms of the $\vec{f}_j$, resultant of
integrating over segments $i$ through $N$. For each $i$, the summation of
\cref{eq:integrated_balance_moment} may be written simply as
\begin{equation}
	\sum\limits_{j=i}^N \vec{I}_j\
\end{equation}
for integrals $\vec{I}_j$. For $j=2,\ldots,N-1$, these are given by
\begin{equation}
	\vec{I}_j = \frac{\ds}{2} \left[\vec{x}_j - \vec{x}_i - \frac{1}{4}\vec{v}_j\right] \cross \vec{f}_j + \frac{\ds}{2} \left[\vec{x}_j - \vec{x}_i - \frac{3}{4}\vec{v}_j\right] \cross \vec{f}_{j+1} \,,
\end{equation}
where $\vec{v}_j = \vec{x}_j - \vec{x}_{j+1}$. For $j=1$, again writing
$d=L(1-e)/(2\ds)$, we have the modified expression
\begin{multline}
	\vec{I}_1 = \frac{\ds}{2}\left[(1+d)(\vec{x}_j-\vec{x}_i) - \frac{1}{4}(1+d)^2\vec{v}_1\right]\cross\vec{f}_1 \\+ \frac{\ds}{2}\left[(1-d)(\vec{x}_j-\vec{x}_i) + \left(\frac{1}{4}(1+d)^2 - 1\right)\vec{v}_1\right]\cross\vec{f}_2\,,
\end{multline}
whilst for $j=N$ we have
\begin{multline}
	\vec{I}_N = \frac{\ds}{2}\left[(1-d)(\vec{x}_j-\vec{x}_i) - \frac{1}{4}(1-d)^2\vec{v}_N\right]\cross\vec{f}_N \\+ \frac{\ds}{2}\left[(1+d)(\vec{x}_j-\vec{x}_i) + \left(\frac{1}{4}(1-d)^2 - 1\right)\vec{v}_N\right]\cross\vec{f}_{N+1}\,.
\end{multline}
These expressions are self-consistent, as taking $d=0$ in the latter two
yields the expression for $\vec{I}_j$.

\section{Integrals as a linear combination}
\label{app:linear_combination}
We decompose the integral of \cref{eq:leading_order_surface} over a straight
segment with endpoints $\vec{x}_j$ and $\vec{x}_{j+1}$, adopting a piecewise
constant discretisation of the force density $\vec{f}$, such that it takes the
value $\vec{f}_j$ on the half of the segment nearest to $\vec{x}_j$, and
$\vec{f}_{j+1}$ otherwise. The limits of integration are determined by
requiring either the coefficient of $\vec{f}_j$ or that of $\vec{f}_{j+1}$,
and for brevity we omit such limits here and will refer instead to the
placeholder $T_{m,q}$ in lieu of $T_{m,q}^L$ and $T_{m,q}^R$ in what follows,
which should be appropriately substituted. Parameterising the straight segment
by $\alpha\in[0,1]$, with $\vec{x}(\alpha)=\vec{x}_j-\alpha\vec{v}$, where
$\vec{v} = \vec{x}_j -
\vec{x}_{j+1}$, and taking $\tensor{K}^{\epsilon}$ to be the kernel of
\cref{eq:leading_order_surface}, we may write the integral over the part of
the segment as
\begin{equation}
\int\tensor{K}^{\epsilon}(\vec{x},s')\vec{f}(s') \intd{s'} =
\tensor{K}^{\epsilon}_I \vec{f}^{\star}\,,
\end{equation}
where $\vec{f}^{\star}$ is the constant force density over the domain of
integration, which is either $\alpha\in[0,1/2]$ or $\alpha\in[1/2,1]$. The
operator $\tensor{K}^{\epsilon}_I$ is given explicitly by
\begin{equation}
	\tensor{K}^{\epsilon}_I = \abs{\vec{v}}\left(\tensor{K}_{\tensor{S}} - \frac{1-e^2}{2e^2}\left[(e^2-s_j'^2)\tensor{K}_{\tensor{D}_0} - 2s_j'\abs{\vec{v}}\tensor{K}_{\tensor{D}_1} - \abs{\vec{v}}^2\tensor{K}_{\tensor{D}_2} \right] \right)\,,
\end{equation}
where the outermost $\abs{\vec{v}}$ term arises due to the change of
integration variable from $s'$ to $\alpha$. In turn, the terms
$\tensor{K}_{\tensor{S}},\tensor{K}_{\tensor{D}_0},\tensor{K}_{\tensor{D}_1},\tensor{K}_{\tensor{D}_2}$
are given by
\begin{align}
	\tensor{K}_{\tensor{S}} &= +\tensor{C}_{0,1}T_{0,-1} + 1(\tensor{C}_{0,3}T_{0,-3} + \tensor{C}_{1,3}T_{1,-3} + \tensor{C}_{2,3}T_{2,-3})\,,\\
	\tensor{K}_{\tensor{D}_0} &= -\tensor{C}_{0,1}T_{0,-3} + 3(\tensor{C}_{0,3}T_{0,-5} + \tensor{C}_{1,3}T_{1,-5} + \tensor{C}_{2,3}T_{2,-5})\,,\\
	\tensor{K}_{\tensor{D}_1} &= -\tensor{C}_{0,1}T_{1,-3} + 3(\tensor{C}_{0,3}T_{1,-5} + \tensor{C}_{1,3}T_{2,-5} + \tensor{C}_{2,3}T_{3,-5})\,,\\
	\tensor{K}_{\tensor{D}_2} &= -\tensor{C}_{0,1}T_{2,-3} + 3(\tensor{C}_{0,3}T_{2,-5} + \tensor{C}_{1,3}T_{3,-5} + \tensor{C}_{2,3}T_{4,-5})\,.
\end{align}
Finally, the coefficients $C_{0,1},C_{0,3},C_{1,3},C_{2,3}$ are determined by
the choice of Taylor expansion point, being either the left or right endpoint
of the segment. When expanding about the left endpoint, where the shifted
rescaled arclength parameter is $s_j'$, we have
\begin{align}
	\tensor{C}_{0,1} &= \tensor{I} \,,\\
	\tensor{C}_{0,3} &= \chi(s_j')\tensor{I} + \vec{w}\vec{w}^T\,,\\
	\tensor{C}_{1,3} &= \abs{\vec{v}}\diff{\chi}{s'}(s_j')\tensor{I} + \vec{w}\vec{v}^T + \vec{v}\vec{w}^T\,,\\
	\tensor{C}_{2,3} &= \frac{1}{2}\abs{\vec{v}}^2\diff{^2\chi}{s'^2}(s_j')\tensor{I} + \vec{v}\vec{v}^T\,,
\end{align}
with $\vec{v}$ as defined previously. Here, $\vec{w}$ joins the evaluation
point to the left endpoint of the segment, which, in the case of
\cref{eq:leading_order_surface}, is given as $\vec{w} = \vec{x}^{S}(s_i',\phi)
- \vec{x}_j$ but may be readily generalised to evaluation points off the
surface of the filament. The corresponding expressions for expansion about the
right endpoint are
\begin{align}
	\tensor{C}_{0,1} &= \tensor{I} \,,\\
	\tensor{C}_{0,3} &= \left(\chi(s_{j+1}') - \abs{\vec{v}}\diff{\chi}{s'}(s_{j+1}') + \frac{1}{2}\abs{\vec{v}}^2\diff{^2\chi}{s'^2}(s_{j+1}')\right)\tensor{I} + \vec{w}\vec{w}^T\,,\\
	\tensor{C}_{1,3} &= \left(\abs{\vec{v}}\diff{\chi}{s'}(s_{j+1}') - \abs{\vec{v}}^2\diff{^2\chi}{s'^2}(s_{j+1}')\right)\tensor{I} + \vec{w}\vec{v}^T + \vec{v}\vec{w}^T\,,\\
	\tensor{C}_{2,3} &= \frac{1}{2}\abs{\vec{v}}^2\diff{^2\chi}{s'^2}(s_{j+1}')\tensor{I} + \vec{v}\vec{v}^T\,.
\end{align}

\section{Explicit antiderivatives}
\label{app:antiderivatives}
Writing $\beta=\beta(\alpha)=B+2C\alpha$ for brevity, the antiderivatives of
$\alpha^mR^q$ for $m=0, q\in\{-1,-3,-5\}$ may be readily computed as
\begin{align}
	\int R^{-1}\intd{\alpha} &= C^{-\frac{1}{2}}\log{\left(\beta + 2C^{\frac{1}{2}}R(\alpha)\right)} \,,\\
	\int R^{-3}\intd{\alpha} &= -\frac{2}{B^2-4AC}\frac{\beta}{R(\alpha)} \,,\\
	\int R^{-5}\intd{\alpha} &= -\frac{2}{3(B^2-4AC)^2}\frac{(B^2-8BC\alpha-4C(3A+2C\alpha^2))\beta}{R(\alpha)^3}\,,
\end{align}
unless we are in the degenerate case, where $B^2-4AC=0$, which yields
\begin{align}
	\int R^{-1}\intd{\alpha} &= C^{-\frac{1}{2}}\sgn{(\beta)}\log{\left(\beta\right)}\,,\\
	\int R^{-3}\intd{\alpha} &= -2C^{\frac{1}{2}}\frac{\sgn{(\beta)}}{\beta^2}\,,\\
	\int R^{-5}\intd{\alpha} &= -4C^{\frac{3}{2}}\frac{\sgn{(\beta)}}{\beta^4}\,.
\end{align}
Here we have assumed that $C>0$, consistent with our assumptions on the
derivatives of $\chi$ and the definition of $C$ in \cref{eq:ABC}. The analysis
of \citet{Walker2020b} and the assumptions of \cref{eq:chi_restrictions} are
sufficient to guarantee that the $R(\alpha)$ is nonzero on $\alpha\in[0,1]$,
thus these integrals are indeed well defined.

\section{Convergence of surface velocity}
\label{app:velocity_convergence}
\begin{figure}
	\centering
	\includegraphics[width=0.7\textwidth]{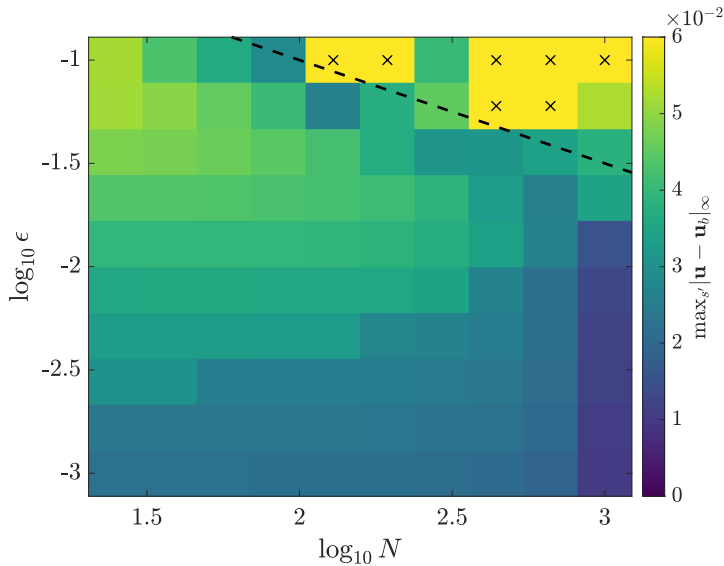}
	\caption{Surface velocity error as a function of slenderness parameter and
	discretisation. We compute the maximum infinity norm error in the surface
	velocity over 2000 points for a straight filament with radius function as
	in \cref{fig:eta}b and a unit background flow $\vec{u}_b=\vec{e}_y$ using
	the method of regularised non-uniform segments. We show in colour the
	error as a function of $\epsilon$ and $N$, with convergence apparent as
	$N$ increases for most common values of $\epsilon$. For both $\epsilon$
	and $N$ large, we see a drastic increase in error, approximately in the
	region bounded below by the black dashed line, which is empirically given
	as $\epsilon\sqrt{N}=1$. Sections marked with a cross exhibit errors
	significantly larger than the range of the colour axis, though these also
	lie outside parameter regimes of typical relevance.}
	\label{fig:app:error}
\end{figure}

For the radius function in \cref{fig:eta}b, we compute the error in the
surface velocity of a straight filament in unit background flow using the RNS
methodology, as in \cref{sec:verify_examples:filament_in_flow} though here
sampling at 2000 points on the surface. The maximum error over the filament
surface is reported in \cref{fig:app:error}, showing substantial refinement as
$N$ increases for common values of slenderness parameter $\epsilon$. Similar
to the method of regularised Stokeslet segments
\citep{Cortez2018,Walker2019d}, for regimes with both large $\epsilon$ and $N$
we see that the error increases dramatically, such that the method is highly
inaccurate, occurring when the parameters are approximately past the threshold
$\epsilon\sqrt{N} = 1$, which is illustrated as a black dashed line in
\cref{fig:app:error}, though this relation is purely empirical. Notably, this
typical breakdown occurs outside regimes of common relevance. Regions marked
with crosses correspond to errors larger than the range of the colour axis.

\begin{thebibliography}{32}
\expandafter\ifx\csname natexlab\endcsname\relax\def\natexlab#1{#1}\fi
\def\au#1{#1} \def\ed#1{#1} \def\yr#1{#1}\def\at#1{#1}\def\jt#1{\textit{#1}}
  \def\bt#1{#1}\def\bvol#1{\textbf{#1}} \def\vol#1{#1} \def\pg#1{#1}
  \def\publ#1{#1}\def\arxiv#1{#1}\def\org#1{#1}\def\st#1{\textit{#1}}

\bibitem[Ainley {\em et~al.\/}(2008)Ainley, Durkin, Embid, Boindala \&
  Cortez]{Ainley2008}
{\sc \au{Ainley, Josephine}, \au{Durkin, Sandra}, \au{Embid, Rafael},
  \au{Boindala, Priya} \& \au{Cortez, Ricardo}} \yr{2008}  \at{{The method of
  images for regularized Stokeslets}}.  \jt{Journal of Computational Physics}
  \bvol{227}~(9),  \pg{4600--4616}.

\bibitem[Chwang \& Wu(1975)]{Chwang1975}
{\sc \au{Chwang, Allen~T.} \& \au{Wu, T. Yao-Tsu}} \yr{1975}
  \at{{Hydromechanics of low-Reynolds-number flow. Part 2. Singularity method
  for Stokes flows}}.  \jt{Journal of Fluid Mechanics}  \bvol{67}~(4),
  \pg{787--815}.

\bibitem[Cortez(2001)]{Cortez2001}
{\sc \au{Cortez, Ricardo}} \yr{2001}  \at{{The method of regularized
  Stokeslets}}.  \jt{SIAM Journal on Scientific Computing}  \bvol{23}~(4),
  \pg{1204--1225}.

\bibitem[Cortez(2018)]{Cortez2018}
{\sc \au{Cortez, Ricardo}} \yr{2018}  \at{{Regularized Stokeslet segments}}.
  \jt{Journal of Computational Physics}  \bvol{375},  \pg{783--796}.

\bibitem[Cortez \& Nicholas(2012)]{Cortez2012}
{\sc \au{Cortez, Ricardo} \& \au{Nicholas, Michael}} \yr{2012}  \at{{Slender
  body theory for Stokes flows with regularized forces}}.  \jt{Communications
  in Applied Mathematics and Computational Science}  \bvol{7}~(1),
  \pg{33--62}.

\bibitem[Cox(1970)]{Cox1970}
{\sc \au{Cox, R.~G.}} \yr{1970}  \at{{The motion of long slender bodies in a
  viscous fluid Part 1. General theory}}.  \jt{Journal of Fluid Mechanics}
  \bvol{44}~(04),  \pg{791--810}.

\bibitem[Curtis {\em et~al.\/}(2012)Curtis, Kirkman-Brown, Connolly \&
  Gaffney]{Curtis2012}
{\sc \au{Curtis, M.~P.}, \au{Kirkman-Brown, J.~C.}, \au{Connolly, T.~J.} \&
  \au{Gaffney, E.~A.}} \yr{2012}  \at{{Modelling a tethered mammalian sperm
  cell undergoing hyperactivation}}.  \jt{Journal of Theoretical Biology}
  \bvol{309},  \pg{1--10}.

\bibitem[Gillies {\em et~al.\/}(2009)Gillies, Cannon, Green \&
  Pacey]{Gillies2009}
{\sc \au{Gillies, Eric~A.}, \au{Cannon, Richard~M.}, \au{Green, Richard~B.} \&
  \au{Pacey, Allan~A.}} \yr{2009}  \at{{Hydrodynamic propulsion of human
  sperm}}.  \jt{Journal of Fluid Mechanics}  \bvol{625},  \pg{445--474}.

\bibitem[Gray(1928)]{Gray1928}
{\sc \au{Gray, James}} \yr{1928} {\em {Ciliary movement}\/}.  \publ{Cambridge
  [England]: Cambridge University Press}.

\bibitem[Gray \& Hancock(1955)]{Gray1955}
{\sc \au{Gray, J.} \& \au{Hancock, G.~J.}} \yr{1955}  \at{{The Propulsion of
  Sea-Urchin Spermatozoa}}.  \jt{Journal of Experimental Biology}
  \bvol{32}~(4),  \pg{802--814}.

\bibitem[Guglielmini {\em et~al.\/}(2012)Guglielmini, Kushwaha, Shaqfeh \&
  Stone]{Guglielmini2012}
{\sc \au{Guglielmini, Laura}, \au{Kushwaha, Amit}, \au{Shaqfeh, Eric S.~G.} \&
  \au{Stone, Howard~A.}} \yr{2012}  \at{{Buckling transitions of an elastic
  filament in a viscous stagnation point flow}}.  \jt{Physics of Fluids}
  \bvol{24}~(12),  \pg{123601}.

\bibitem[Hall-Mcnair {\em et~al.\/}(2019)Hall-Mcnair, Montenegro-Johnson,
  Gad{\^{e}}lha, Smith \& Gallagher]{Hall-McNair2019}
{\sc \au{Hall-Mcnair, Atticus~L.}, \au{Montenegro-Johnson, Thomas~D.},
  \au{Gad{\^{e}}lha, Hermes.}, \au{Smith, David~J.} \& \au{Gallagher,
  Meurig~T.}} \yr{2019}  \at{{Efficient implementation of elastohydrodynamics
  via integral operators}}.  \jt{Physical Review Fluids}  \bvol{4}~(11),
  \pg{1--24}.

\bibitem[Hancock(1953)]{Hancock1953}
{\sc \au{Hancock, G.~J.}} \yr{1953}  \at{{The self-propulsion of microscopic
  organisms through liquids}}.  \jt{Proceedings of the Royal Society of London.
  Series A. Mathematical and Physical Sciences}  \bvol{217}~(1128),
  \pg{96--121}.

\bibitem[Ishimoto \& Gaffney(2018)]{Ishimoto2018a}
{\sc \au{Ishimoto, Kenta} \& \au{Gaffney, Eamonn~A.}} \yr{2018}  \at{{An
  elastohydrodynamical simulation study of filament and spermatozoan swimming
  driven by internal couples}}.  \jt{IMA Journal of Applied Mathematics}
  \bvol{83}~(4),  \pg{655--679}.

\bibitem[Johnson(1980)]{Johnson1980}
{\sc \au{Johnson, Robert~E.}} \yr{1980}  \at{{An improved slender-body theory
  for Stokes flow}}.  \jt{Journal of Fluid Mechanics}  \bvol{99}~(2),
  \pg{411--431}.

\bibitem[Keller \& Rubinow(1976)]{keller1976}
{\sc \au{Keller, Joseph~B} \& \au{Rubinow, Sol~I}} \yr{1976}  \at{{Slender-body
  theory for slow viscous flow}}.  \jt{Journal of Fluid Mechanics}
  \bvol{75}~(4),  \pg{705--714}.

\bibitem[Lighthill(1976)]{lighthill1976}
{\sc \au{Lighthill, James}} \yr{1976}  \at{{Flagellar hydrodynamics}}.
  \jt{SIAM review}  \bvol{18}~(2),  \pg{161--230}.

\bibitem[Moreau {\em et~al.\/}(2018)Moreau, Giraldi \&
  Gad{\^{e}}lha]{Moreau2018}
{\sc \au{Moreau, Cl{\'{e}}ment}, \au{Giraldi, Laetitia} \& \au{Gad{\^{e}}lha,
  Hermes}} \yr{2018}  \at{{The asymptotic coarse-graining formulation of
  slender-rods, bio-filaments and flagella}}.  \jt{Journal of The Royal Society
  Interface}  \bvol{15}~(144),  \pg{20180235}.

\bibitem[Neal {\em et~al.\/}(2020)Neal, Hall-McNair, Kirkman-Brown, Smith \&
  Gallagher]{Neal2020}
{\sc \au{Neal, Cara~V.}, \au{Hall-McNair, Atticus~L.}, \au{Kirkman-Brown,
  Jackson}, \au{Smith, David~J.} \& \au{Gallagher, Meurig~T.}} \yr{2020}
  \at{{Doing more with less: The flagellar end piece enhances the propulsive
  effectiveness of human spermatozoa}}.  \jt{Physical Review Fluids}
  \bvol{5}~(7),  \pg{073101}.

\bibitem[Olson {\em et~al.\/}(2013)Olson, Lim \& Cortez]{Olson2013}
{\sc \au{Olson, Sarah~D.}, \au{Lim, Sookkyung} \& \au{Cortez, Ricardo}}
  \yr{2013}  \at{{Modeling the dynamics of an elastic rod with intrinsic
  curvature and twist using a regularized Stokes formulation}}.  \jt{Journal of
  Computational Physics}  \bvol{238},  \pg{169--187}.

\bibitem[Pozrikidis(1992)]{pozrikidis1992}
{\sc \au{Pozrikidis, Constantine}} \yr{1992} {\em {Boundary integral and
  singularity methods for linearized viscous flow}\/}.  \publ{Cambridge
  University Press}.

\bibitem[Pozrikidis(2010)]{Pozrikidis2010}
{\sc \au{Pozrikidis, C.}} \yr{2010}  \at{{Shear flow over cylindrical rods
  attached to a substrate}}.  \jt{Journal of Fluids and Structures}
  \bvol{26}~(3),  \pg{393--405}.

\bibitem[Roper {\em et~al.\/}(2006)Roper, Dreyfus, Baudry, Fermigier, Bibette
  \& Stone]{Roper2006}
{\sc \au{Roper, Marcus}, \au{Dreyfus, R{\'{e}}mi}, \au{Baudry, Jean},
  \au{Fermigier, M.}, \au{Bibette, J.} \& \au{Stone, H.~A.}} \yr{2006}  \at{{On
  the dynamics of magnetically driven elastic filaments}}.  \jt{Journal of
  Fluid Mechanics}  \bvol{554},  \pg{167--190}.

\bibitem[du~Roure {\em et~al.\/}(2019)du~Roure, Lindner, Nazockdast \&
  Shelley]{DuRoure2019}
{\sc \au{du~Roure, Olivia}, \au{Lindner, Anke}, \au{Nazockdast, Ehssan~N.} \&
  \au{Shelley, Michael~J.}} \yr{2019}  \at{{Dynamics of Flexible Fibers in
  Viscous Flows and Fluids}}.  \jt{Annual Review of Fluid Mechanics}
  \bvol{51}~(1),  \pg{539--572}.

\bibitem[Schoeller \& Keaveny(2018)]{Schoeller2018}
{\sc \au{Schoeller, Simon~F.} \& \au{Keaveny, Eric~E.}} \yr{2018}  \at{{From
  flagellar undulations to collective Motion: Predicting the dynamics of sperm
  suspensions}}.  \jt{Journal of the Royal Society Interface}  \bvol{15}~(140),
   \arxiv{arXiv: 1801.08180}.

\bibitem[Shampine \& Reichelt(1997)]{Shampine1997}
{\sc \au{Shampine, Lawrence~F.} \& \au{Reichelt, Mark~W.}} \yr{1997}  \at{{The
  MATLAB ODE Suite}}.  \jt{SIAM Journal on Scientific Computing}
  \bvol{18}~(1),  \pg{1--22}.

\bibitem[Simons {\em et~al.\/}(2015)Simons, Fauci \& Cortez]{Simons2015}
{\sc \au{Simons, Julie}, \au{Fauci, Lisa} \& \au{Cortez, Ricardo}} \yr{2015}
  \at{{A fully three-dimensional model of the interaction of driven elastic
  filaments in a Stokes flow with applications to sperm motility}}.
  \jt{Journal of Biomechanics}  \bvol{48}~(9),  \pg{1639--1651}.

\bibitem[Smith(2009)]{Smith2009d}
{\sc \au{Smith, D.~J.}} \yr{2009}  \at{{A boundary element regularized
  Stokeslet method applied to cilia- and flagella-driven flow}}.
  \jt{Proceedings of the Royal Society A: Mathematical, Physical and
  Engineering Sciences}  \bvol{465}~(2112),  \pg{3605--3626},  \arxiv{arXiv:
  1008.0570}.

\bibitem[Smith {\em et~al.\/}(2019)Smith, Montenegro-Johnson \&
  Lopes]{Smith2019}
{\sc \au{Smith, David~J.}, \au{Montenegro-Johnson, Thomas~D.} \& \au{Lopes,
  Susana~S.}} \yr{2019}  \at{{Symmetry-Breaking Cilia-Driven Flow in
  Embryogenesis}}.  \jt{Annual Review of Fluid Mechanics}  \bvol{51}~(1),
  \pg{105--128}.

\bibitem[Walker {\em et~al.\/}(2020)Walker, Curtis, Ishimoto \&
  Gaffney]{Walker2020b}
{\sc \au{Walker, Benjamin~J.}, \au{Curtis, Mark~P.}, \au{Ishimoto, Kenta} \&
  \au{Gaffney, Eamonn~A.}} \yr{2020}  \at{{A regularised slender-body theory of
  non-uniform filaments}}.  \jt{Journal of Fluid Mechanics}  \bvol{899},
  \pg{A3}.

\bibitem[Walker {\em et~al.\/}(2019{\natexlab{{\em a\/}}})Walker, Ishimoto,
  Gad{\^{e}}lha \& Gaffney]{Walker2019d}
{\sc \au{Walker, Benjamin~J.}, \au{Ishimoto, Kenta}, \au{Gad{\^{e}}lha, Hermes}
  \& \au{Gaffney, Eamonn~A.}} \yr{2019{\natexlab{{\em a\/}}}}  \at{{Filament
  mechanics in a half-space via regularised Stokeslet segments}}.  \jt{Journal
  of Fluid Mechanics}  \bvol{879},  \pg{808--833}.

\bibitem[Walker {\em et~al.\/}(2019{\natexlab{{\em b\/}}})Walker, Ishimoto \&
  Gaffney]{Walker2019g}
{\sc \au{Walker, Benjamin~J.}, \au{Ishimoto, Kenta} \& \au{Gaffney, Eamonn~A.}}
  \yr{2019{\natexlab{{\em b\/}}}}  \at{{A new basis for filament simulation in
  three dimensions}} ,  \arxiv{arXiv: 1907.04823}.

\end{thebibliography}
\end{document}